\documentclass{article}
\usepackage{amsmath,amsfonts,epsfig,lscape} 
\usepackage{float}
\makeatletter
\let\original@start@gather=\start@gather
\def\start@gather{\everydisplay{}\original@start@gather}
\makeatother

\title{Aperture Array Configurations for SKA1 Core}

\begin{document}
\large
\noindent

\maketitle

%\title{Aperture Array Configurations for SKA1 Core}
\begin{center}
Keith Grainge
\end{center}

\vspace{5mm}

\section{Introduction}

This memo considers some aspects of the configuration of the SKA1 Low
Frequency Aperture Array, both at the element and station level. At
the element level I propose a possible scenario for forming station
beams where elements are shared between stations and apodisation is
implemented, with the aim of improving filling factor, overall
sensitivity and sidelobe performance; the disadvantages of such a
scheme with regards to beam former requirements and shortest available
baseline are also discussed. At the station level, a randomised
configuration within a filled central region together with spiral arms
is explored.

\subsection{Configuration of elements}

\begin{figure}[H]
\begin{center}
\includegraphics[width=8cm,angle=0]{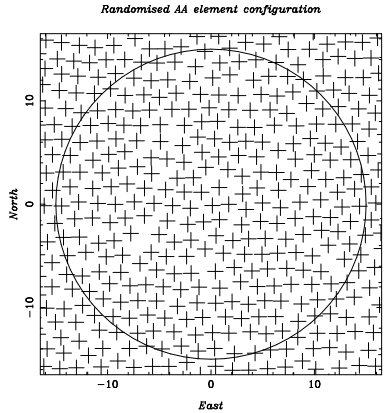}
\caption{\label{element_config} A 35m diameter station comprised of
  256 elements arranged in a randomly scattered sea of log-periodic
  antennas.}
\end{center}
\end{figure}

In this memo I assume that the elements will be placed in a randomly
scattered configuration subject to the constraints of a minimum
distance between elements, $d_{min}$, and that the desired array
filling factor, $f_{fill}$ be achieved (see
Figure~\ref{element_config}). This is the preferred element placement
scheme \cite{eloy} proposed by the designers of the log-periodic
antennas \cite{eloy2} that have been adopted in the SKA Baseline
Design\cite{BD}.

\section{Shared element station configuration}\label{shared}

In this section I consider the possibility that the signals from
individual elements could contribute to several station beams. This
also allows for the possibility of implementing apodisation of the
station beam without the loss in overall sensitivity usually
associated with apodisation.

\begin{figure}[H]
\begin{center}
\includegraphics[width=8cm,angle=0]{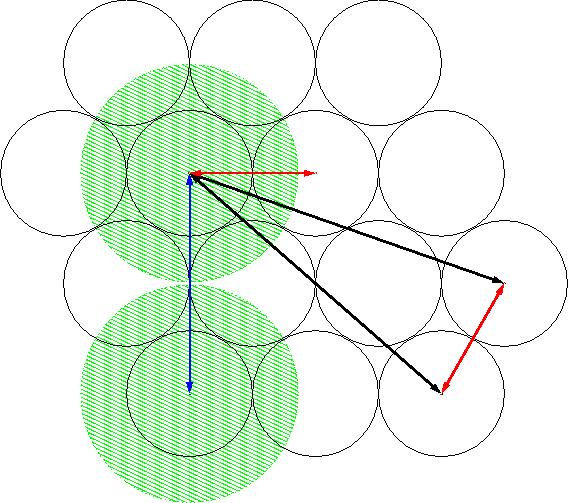}
\caption{\label{hex_pack}Hexagonal close packed configuration for
  station centres. The black circles have a radius, $r_{full}$, equal
  to half the distance between adjacent station centres and all the
  elements within this area receive their full weight in the beam
  formation process. The green filled circles have radius equal to
  $\sqrt{3}r_{full}$ and indicated the full extent of the area over
  which elements will contribute to this particular station beam. Red
  arrows shows one of the baselines between adjacent stations which
  will be discarded before imaging. The shortest legitimate baseline
  occurs between stations whose green circles are adjacent to each
  other and do not share any elements and the blue line shows as
  example of these.  The black baselines are both examples of
  legitimate baselines --- note that both of the baselines between one
  station and a pair of adjacent stations are legitimate, despite the
  fact that the baseline between these two is discarded. }
\end{center}
\end{figure}

The inner $\approx 250$m radius of the core of SKA-low is required to be as
close to fully filled as is possible; I therefore consider a
hexagonal close packed arrangement of stations centres as shown in
Figure~\ref{hex_pack}.
The scale of hexagonal close packed configuration is defined by the
radius of the circles of which it is comprised, called $r_{full}$ for
reasons that will become apparent later.

For sake of comparison, I first consider a ``standard'' set up in
which stations comprise elements within $r_{full}$ of the station
centre, which are all uniformly weighted during beam forming. The
problems with such a solution include:
\begin{itemize}
\item The overall filling factor of the aperture array will be reduced by
\item There are areas between the stations that are not used by any
  stations. It would be wasteful to deploy elements in these areas
  given that they would not be used, but the existence of these
  ``voids'' has the following implications:
\begin{itemize}
\item It will be impossible to implement the flexible station size
  option proposed by the LFAA consortium \cite{LFAA} since these stations
  would contain void areas not populated by elements resulting in a
  very poor station beam. Note that for the same reason beam forming
  large areas for ionospheric calibration (see e.g. \cite{stefan}) would
  also be problematic.
\item \cite{eloy3} find that the beam pattern for individual elements
  taking into account cross-coupling can be well approximated by
  simulating the response assuming that the elements sit in a
  semi-infinite sea of elements. Elements located at the edge of
  stations adjacent to voids will have a significantly different
  response and this will lead to a poorly known overall station beam.
\end{itemize}
\item The sharp cut off in station illumination will result in a beam
  pattern will have very strong sidelobes, with strong implications
  for the possible dynamic range achievable by the telescope (see
  e.g. \cite{braun} for further discussion).
\end{itemize}

\begin{figure}[H]
\begin{center}
\includegraphics[width=8cm,angle=0]{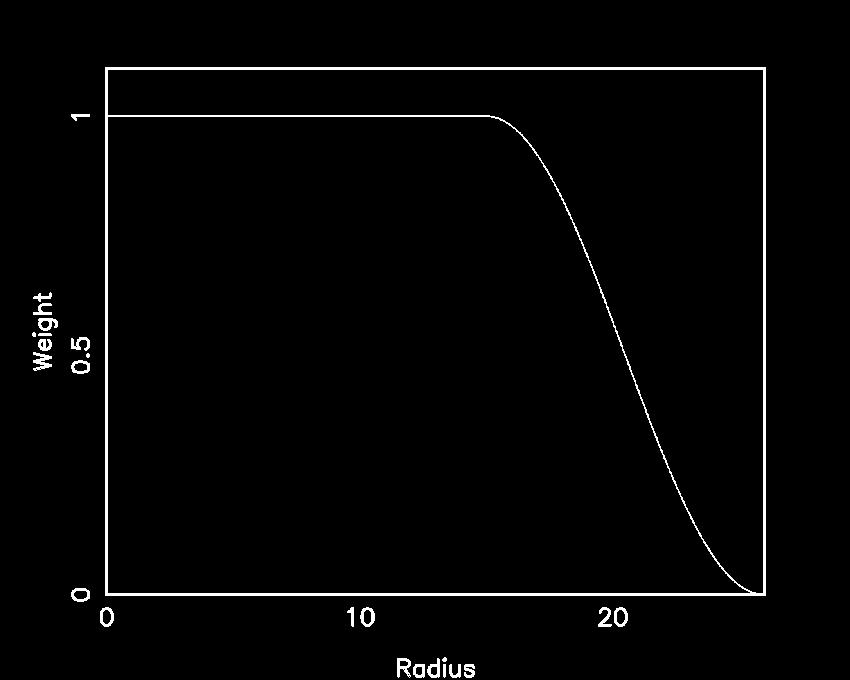}
\caption{\label{rad_weight}Hanning tapered radial station weighting
  function (see eqn~\ref{weight_fn}).}
\end{center}
\end{figure}

\begin{figure}[H]
\begin{center}
\includegraphics[width=8cm,angle=0]{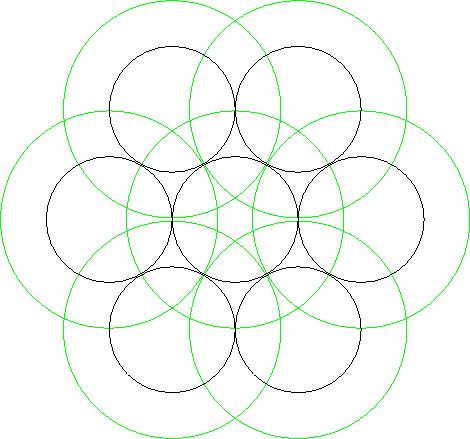}
\caption{\label{overlap}Diagram showing the degree of overlap between
  stations and that individual elements will therefore be included in multiple
  different station beams}
\end{center}
\end{figure}

By contrast, in the shared-element configuration, stations are formed
from elements within a larger radius greater, and here I consider a
scenario in which all those elements with $\sqrt{3}r_{full}$ will
contribute. However, those within $r_{full}$ will be added with full
weight during beam formation while those beyond will be apodised with
a Hanning window function (see Figure~\ref{rad_weight}) i.e.
\begin{alignat}{2}\label{weight_fn}
w & = 1 & \quad  &r\le r_{full}\\
w & =
\frac{1}{2}\left(1+\cos{\left(\frac{\pi\left(r-r_{full}\right)}{\left(\sqrt{3}-1\right)r_{full}}\right)}\right)
\quad& r_{full}\le &r\le \sqrt{3}r_{full} \\
\end{alignat}
Therefore the signals from individual elements will be included in
multiple different station beams (see Figure~\ref{overlap}); on
average they will appear in $3\sqrt{3}/2$. There are no voids in this
configuration, so all the problems with the uniformly weighted
configuration disappear.

The possibility therefore exists that visibilities could be formed
between stations which incorporate common elements in common. This
would give rise to autocorrelation type terms appearing in the
visibilities; this is undesirable.  Therefore any visibilities formed
from stations where the baseline length is less than
$2\sqrt{3}r_{full}$ must be discarded --- examples of these baselines
appear in red and only occur between adjacent stations in the
hexagonal close packed configuration.  Approximately $3N$ of the
$N(N-1)/2$ baselines between the $N$ stations are discarded for this
reason.  The shortest remaining baselines (hereafter referred to as
``legitimate'' baselines) occur between stations whose total extent
(indicated by the green circles) are adjacent to each other and do not
share any elements.  However, both of the baselines between one
station and a pair of adjacent stations are legitimate, despite the
fact that the baseline between these two is discarded.

\subsection{Benefits}
\begin{figure}[H]
\begin{center}
\includegraphics[width=5cm,angle=0]{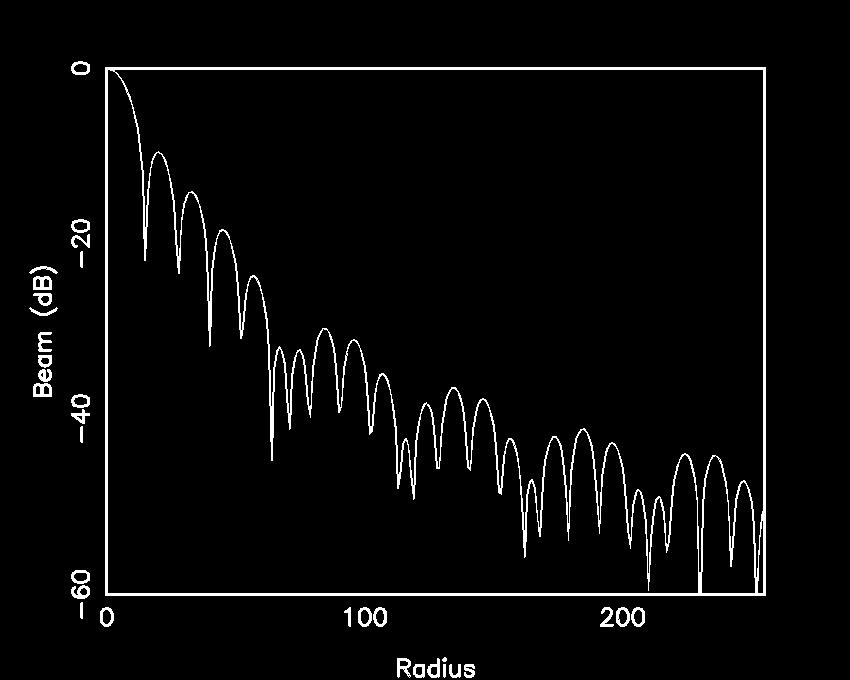}
\includegraphics[width=5cm,angle=0]{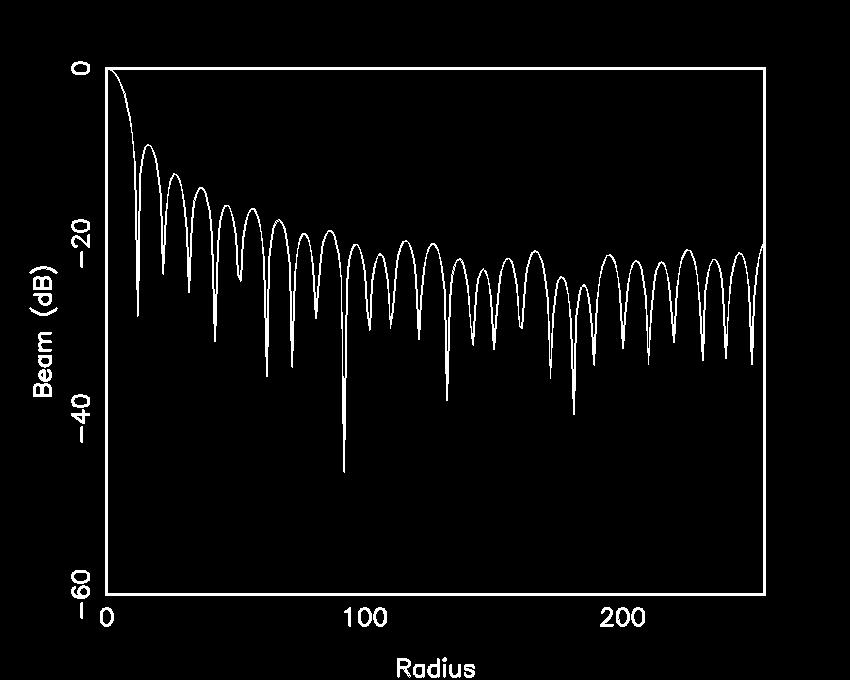}
\caption{\label{sidelobe}Plot which shows on the left the beam pattern
  for a (filled) station using the weighting function in
  eqn~\ref{weight_fn}; and on the right a uniformly weighted circular
  station.}
\end{center}
\end{figure}

\begin{itemize}
\item The tapering function greatly improves the station sidelobes,
  which will therefore suppress the effects of bright sources far from
  the field centre and so in turn improve imaging and calibration, see
  Figure~\ref{sidelobe}.
\item With no voids, the element filling factor in the core is maximised and
  depends only upon the inter-element spacing.
\item A flexible station size now becomes feasible; even if this is
  not chosen for Phase 1 of the SKA, a flexible station configuration
  is an attractive upgrade path as beam former technology becomes
  cheaper, but a configuration with voids will prevent this from being
  possible to implement.
\end{itemize}

\subsection{Disadvantages}
\begin{itemize}
\item The beamformer will be more complicated: element signals will be
  required by more than one station beamformer and the total number of
  data points to be summed will increase by $3\sqrt{3}/2$.
\item The shortest baseline, which determines the maximum angular
  scale sampled by the telescope, will be $\sqrt{3}$ times longer than
  that achieved by a configuration where elements are not shared.
\end{itemize}

\subsection{Noise considerations}

\begin{figure}[H]
\begin{center}
\includegraphics[width=10cm,angle=0]{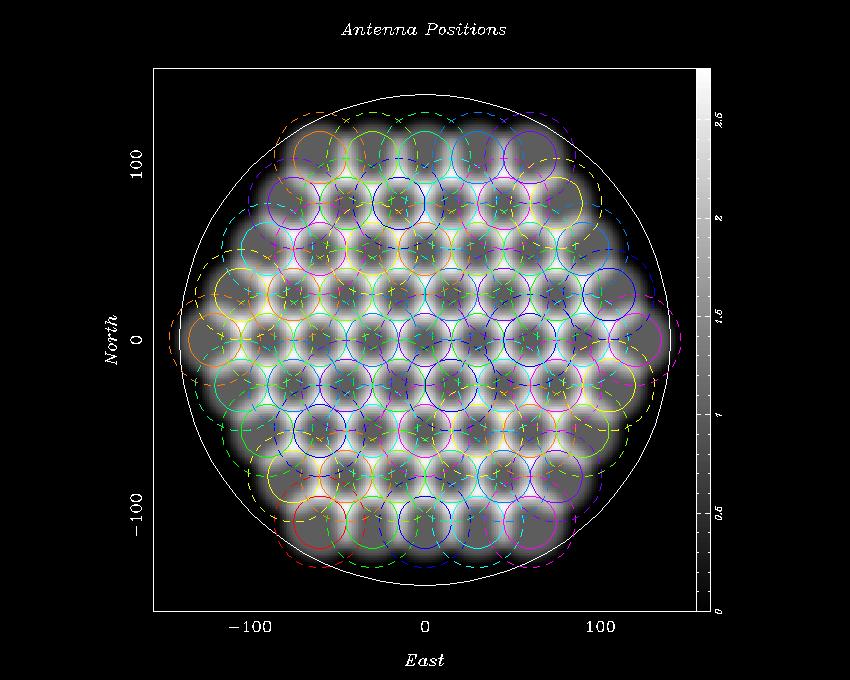}
\caption{\label{weight}Plot showing the sum  weights for a
  hexagonally close packed configuration of stations using the
  individual station weights according to Eqn~\ref{weight_fn}. The
  regions of very high weight arise from regions where 3 stations
  contribute with just short of full individual weight.}
\end{center}
\end{figure}

Compared to a telescope with circular stations of radius $r_{full}$,
the proposed configuration has much the same number of baselines
(apart from those that are discarded) but a significantly larger
collecting area per station; so it is tempting to conclude that this
implies an improvement in the sensitivity, but this is not the
case. The reason for this is that the visibilities formed from
baselines such as the black ones those shown in Figure~\ref{hex_pack}
(hereafter known as ``non-independent'' baselines) are significantly
correlated and so the noise level will not decrease as $\sqrt
n_{baselines}$; this is discussed further in
Section~\ref{correlations}. There will in fact be a degradation in the
maximally achievable sensitivity resulting from the fact that some
elements receive more weight than others due to their position. The
extent to which this is a problem can be judged from the summed
weights, which should ideally be uniform; see Figure~\ref{weight}.

\subsection{Visibility correlations}\label{correlations}

\begin{figure}[H]
\begin{center}
\includegraphics[width=5cm,angle=0]{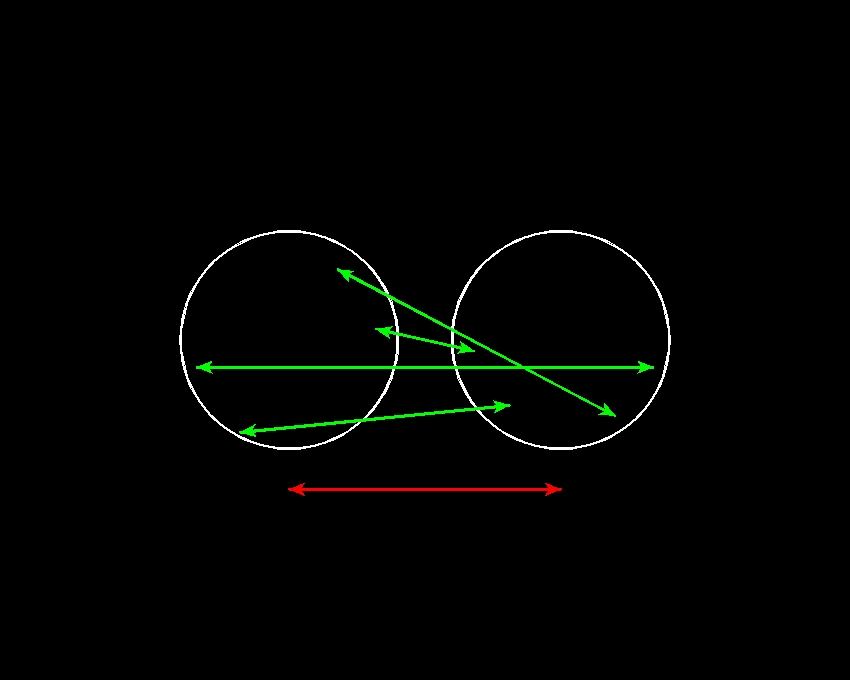}\qquad
\includegraphics[width=5cm,angle=0]{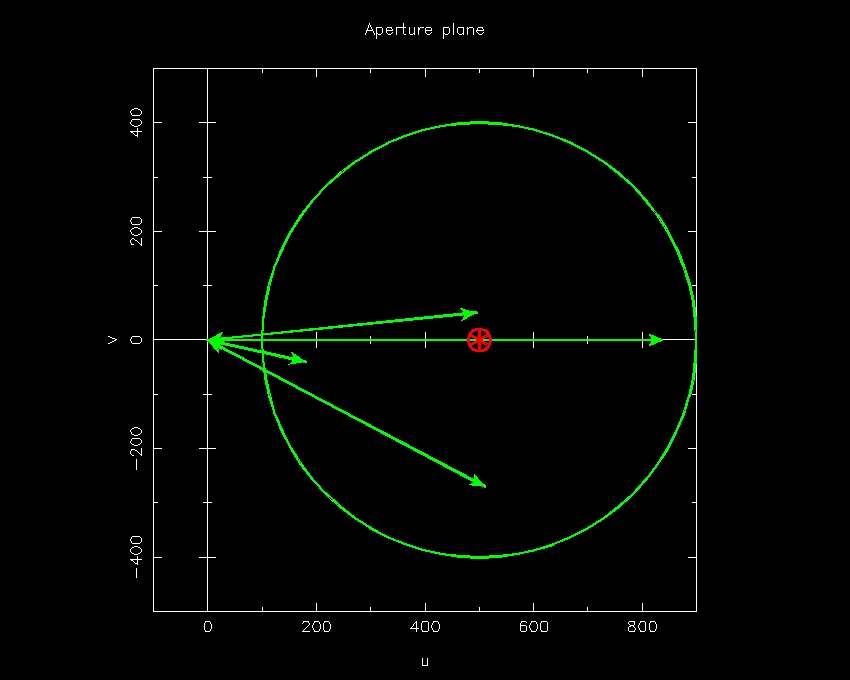}
\caption{\label{indiv_basel}The lefthand plot shows 2 stations and
  some of the individual baselines that could be formed from
  correlating elements within each. These baselines are then also
  shown in the righthand plot in the {\it uv}-plane together with the
  extent of the aperture illumination function for the correlation of
  the two stations. The redarrow shows the baseline between the
  centres of the two stations and the redpoint show the corresponding
  {\it uv} position in the aperture plane.}
\end{center}
\end{figure}

As stated above, visibilities from non-independent baselines such as
the black ones those shown in Figure~\ref{hex_pack} are significantly
correlated. However, this is not necessarily a problem. It is always
the case that two visibilities will be correlated when there is
overlap between their {\it uv}-positions on being convolved by the
aperture illumination function, because they are sampling a common
region of the underlying {\it uv}-plane. This correlation between the
signals must be accounted for in the covariance matrix when analysing
the data in the {\it uv}-plane, but the thermal noise is usually
assumed to be diagonal. However, for these non-independent baselines,
there will also be a correlated component to their noises, which must
included in the covariance matrix. This idea is further illustrated in
Figure~\ref{indiv_basel}. The signal from each station can be
considered to be the sum of the voltages from each of its comprising
elements. Therefore the visibility formed by the correlation of the
two stations is the sum of all the possible products between the two
stations. The visibility is therefore a weighted sum of the values in
the {\it uv}-plane within a patch defined by the convolution of the
two station areas (i.e. the aperture illumination function). The
non-independent baselines will therefore share some of these product
components between individual elements. The result will be that these
components will be upweighted with respect to the rest. This is
therefore equivalent in standard interferometry to arbitrarily
upweighting certain visibilities in the {\it uv}-plane and will
therefore lead to a sub-optimal signal-to-noise but will not introduce
biases.
 
\section{Modifications to the shared element configuration}

\subsection{Randomised station positions}\label{random_stn}

Using a hexagonal close packed configuration gives rise to a large
number of redundant baselines, which will be poor for filling of the
{\it uv}-plane and hence for imaging. Also, since these redundant
baselines will have somewhat different station beam patterns due to
the random distribution of elements, the level of redundancy is
unlikely to be useful for calibration. Therefore a certain amount of
reduction in filling factor and a randomisation of the station centres
may be valuable.

\subsection{Different apodisation functions}

\begin{figure}[H]
\begin{center}
\includegraphics[width=5cm,angle=0]{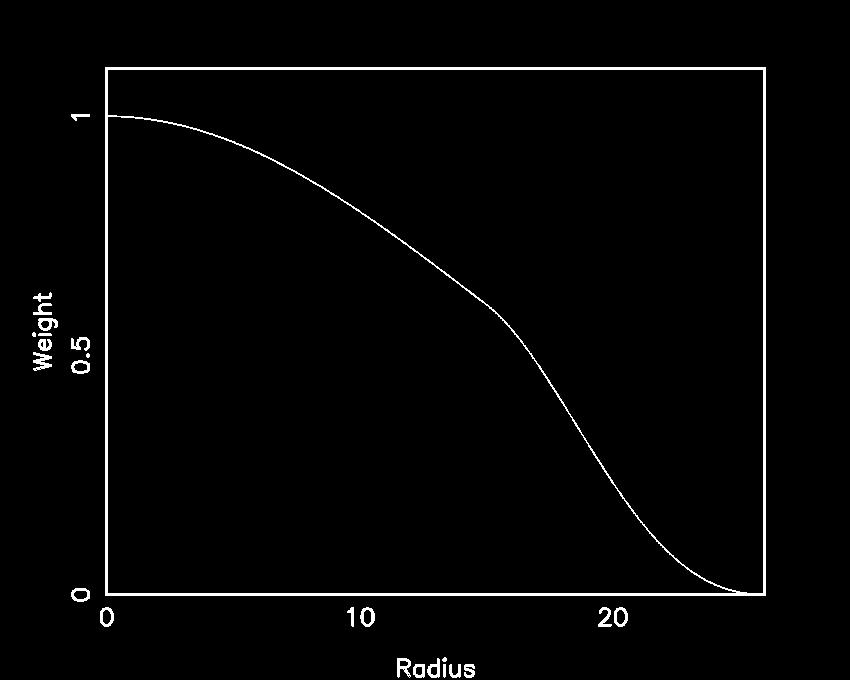}\qquad
\includegraphics[width=5cm,angle=0]{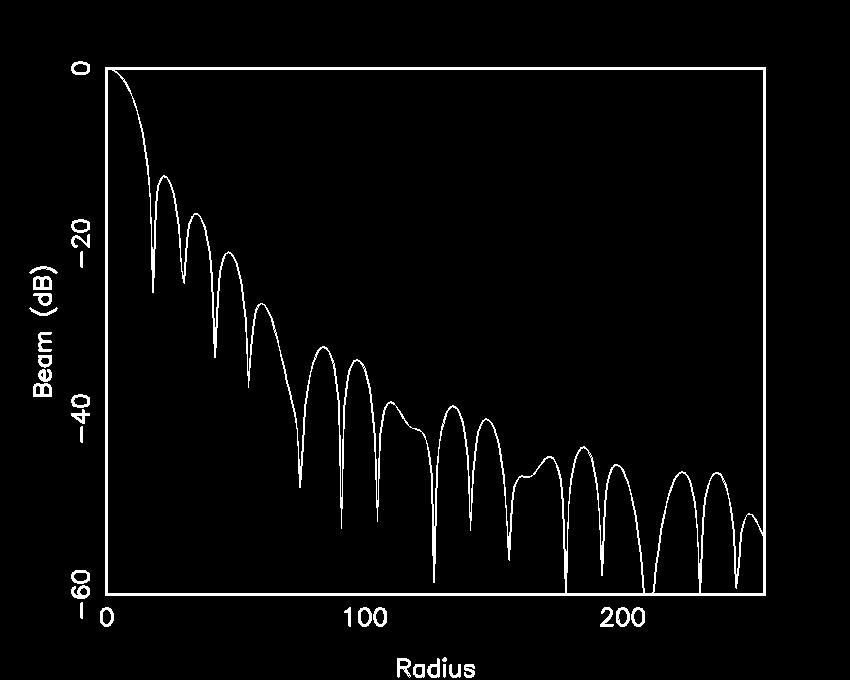}\\
\includegraphics[width=10cm,angle=0]{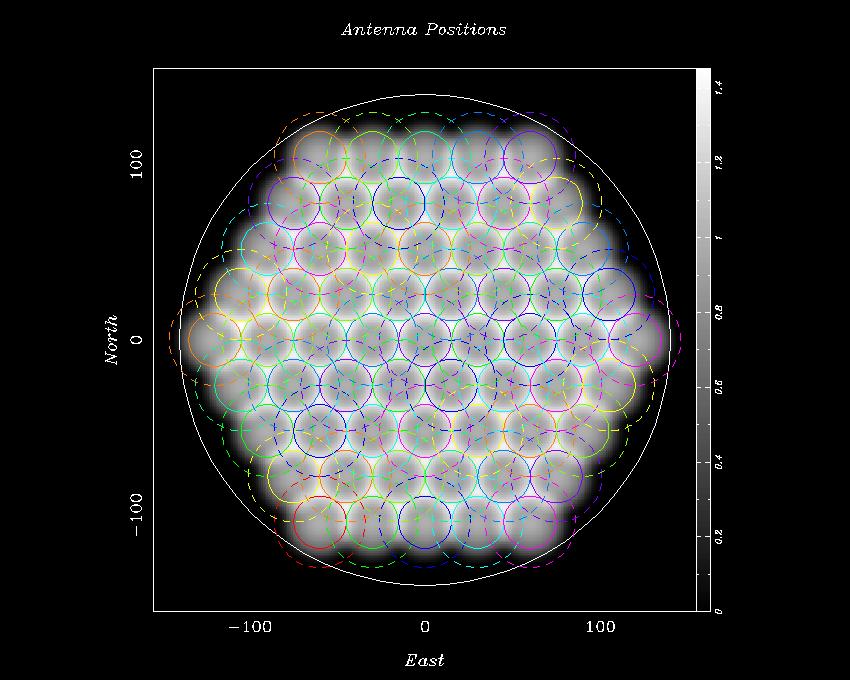}\\
\caption{\label{gau}Plots for a tapered Gaussian station weighting
  function. From top left: the radial weighting function itself; the
  associated beam response; the sum weights for a hexagonally close
  packed configuration of stations.}
\end{center}
\end{figure}

Figure~\ref{weight} shows that some elements are significantly
upweighted with respect the mean which will degrade the overall
sensitivity of the telescope. Randomising the station centres may
alleviate this problem to a certain extent. However, adopting a
different weighting function should also be considered; plots for a
tapered Gaussian weight function of the same extent are shown in
Figure~\ref{gau}. The results look attractive but an optimising the
weighting function has not been attempted in this paper.

Another possible way forward would be to extend the extent of the
tapered region for each station, for example having a Hanning window
out to say $2r_{full}$. The disadvantages of implementing this are:
\begin{itemize}
\item Increased complexity for the beam former.
\item Increased number of baselines that must be rejected since the
  component stations share elements.
\end{itemize}

One final possibility that should be considered is whether one needs a
``guard area'' between adjacent stations whose signals are correlated,
since there will certainly be cross-coupling between adjacent elements
leading to auto-correlation terms appearing in the visibilities (even
if these are significantly downweighted).

\subsection{Apodisation through thinning}

Apodisation can be achieved either through downweighting of outlying
elements in an element or through reducing the density of the elements
as a function of radius (i.e. thinning) and weighting them all
equally. Downweighting of outlying elements will produce the better
beam pattern, but outside the core it will be wasteful of element
collecting area, so thinning may be a good compromise; it will
certainly give a better beam pattern than a uniformly weighted,
uniformly dense, circular station.

\section{Limits to the utility of apodisation}

Apodisation certainly improves the inner regions of the station beam,
but some far-out sidelobes are not suppressed by this method. The
reason for this can be seen through an understanding of what generates
the various features in the station beam and is argued below in a
somewhat pedagogical fashion; experts may wish to skip to
Section~\ref{stn_conf}.

\subsection{Understanding the station beam of an aperture array}

\begin{landscape}

\begin{figure}[H]
\includegraphics[width=6.75cm]{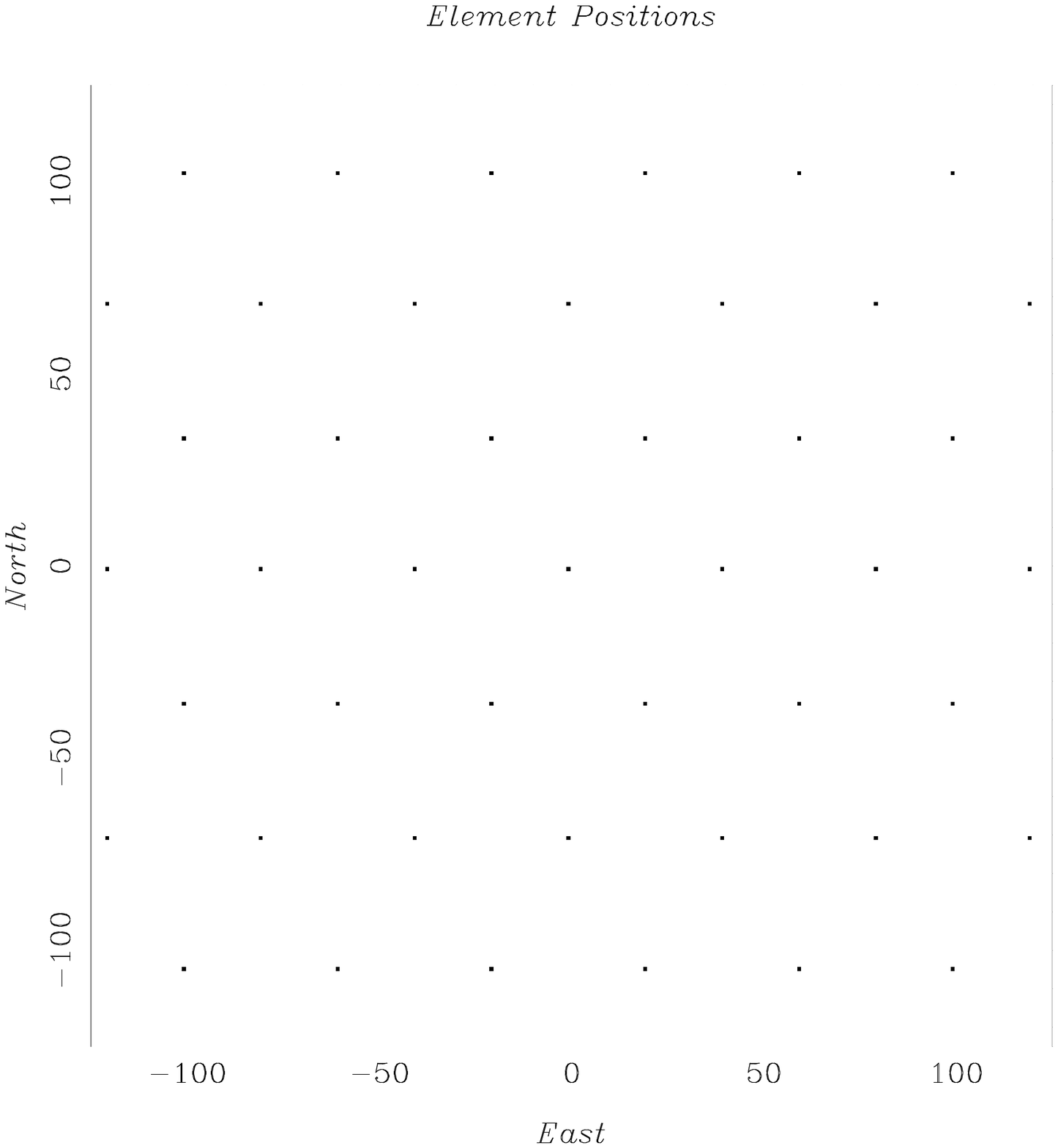}\includegraphics[width=6.75cm]{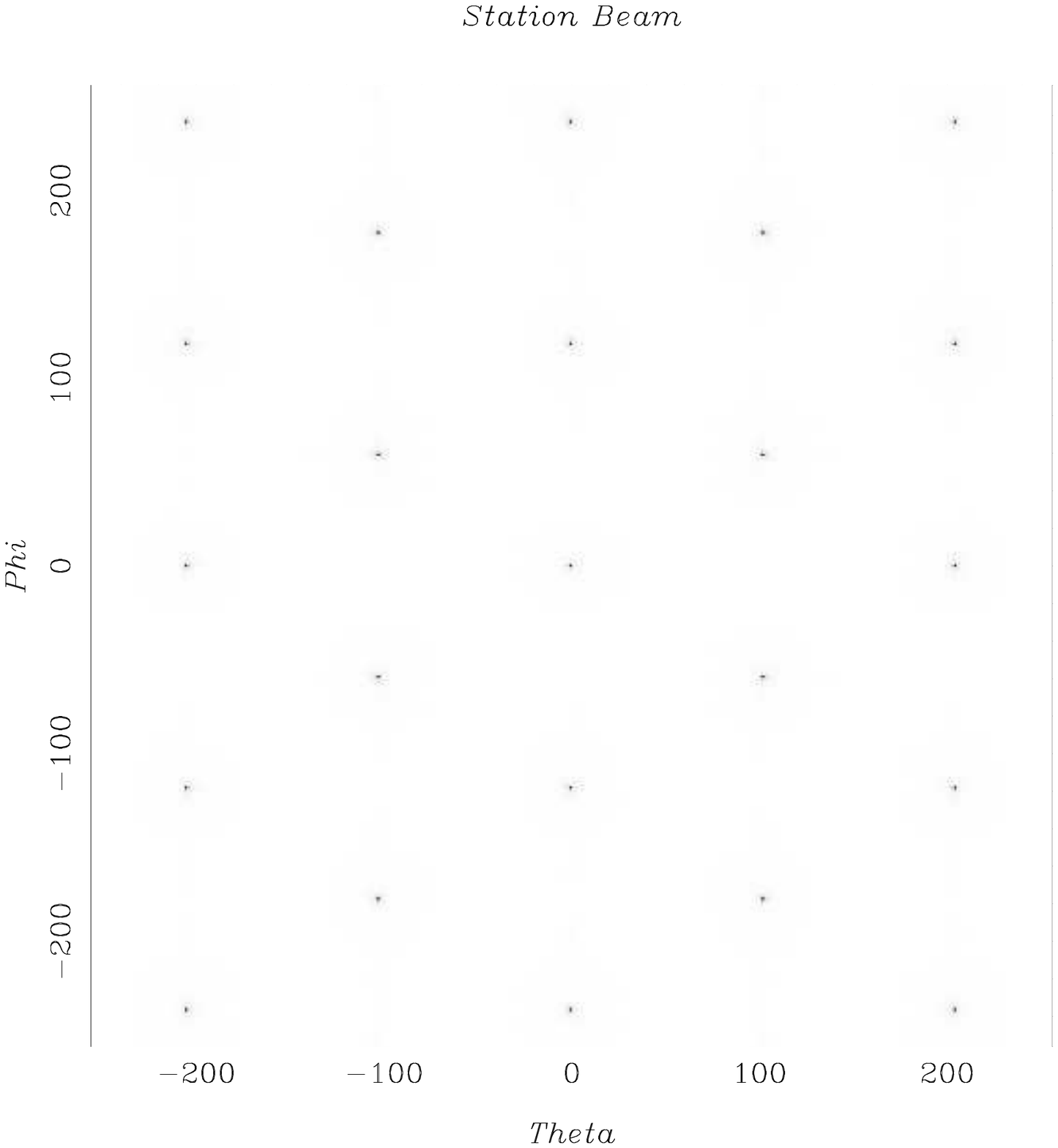}\quad\includegraphics[width=6cm]{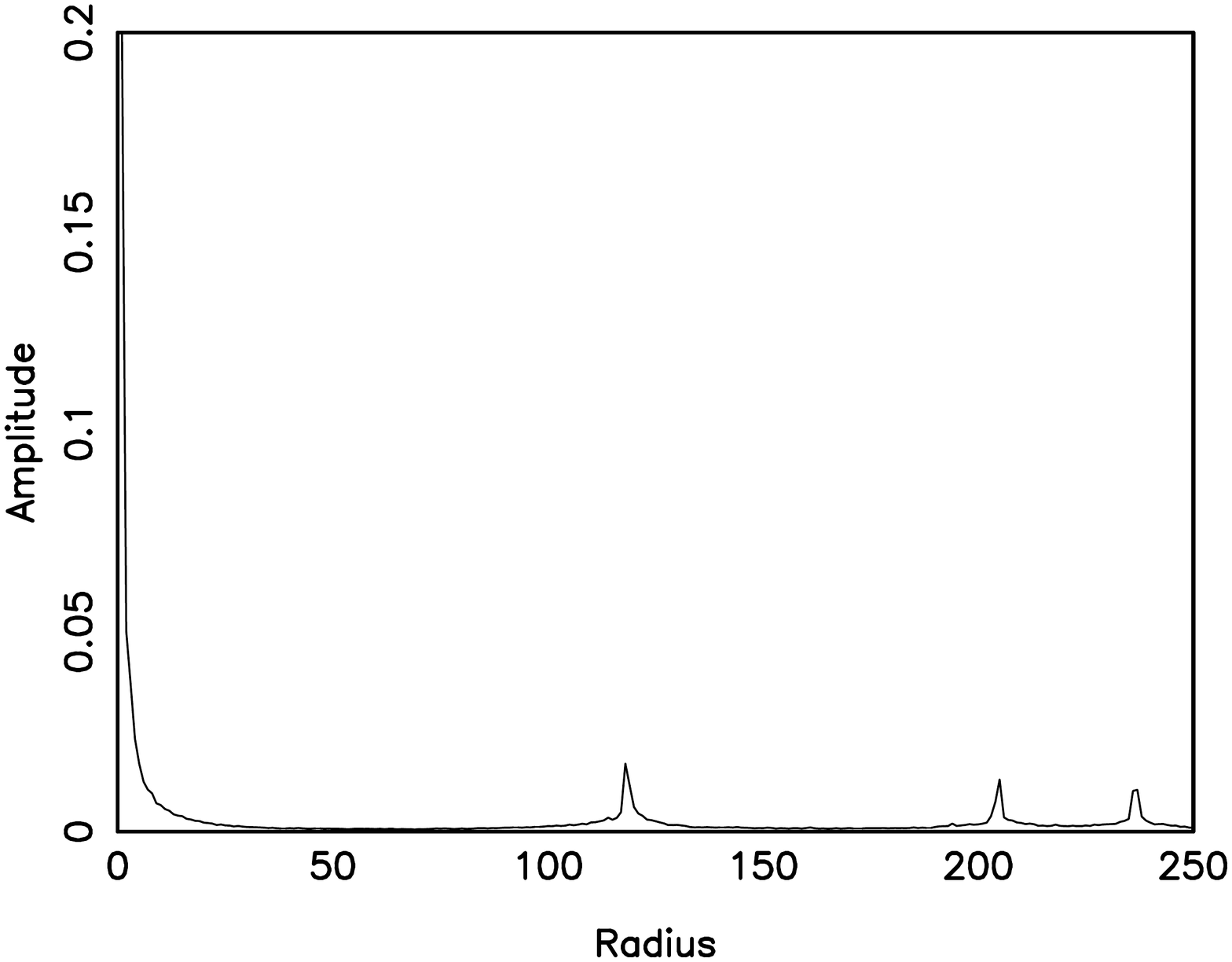}\\
\includegraphics[width=6.75cm]{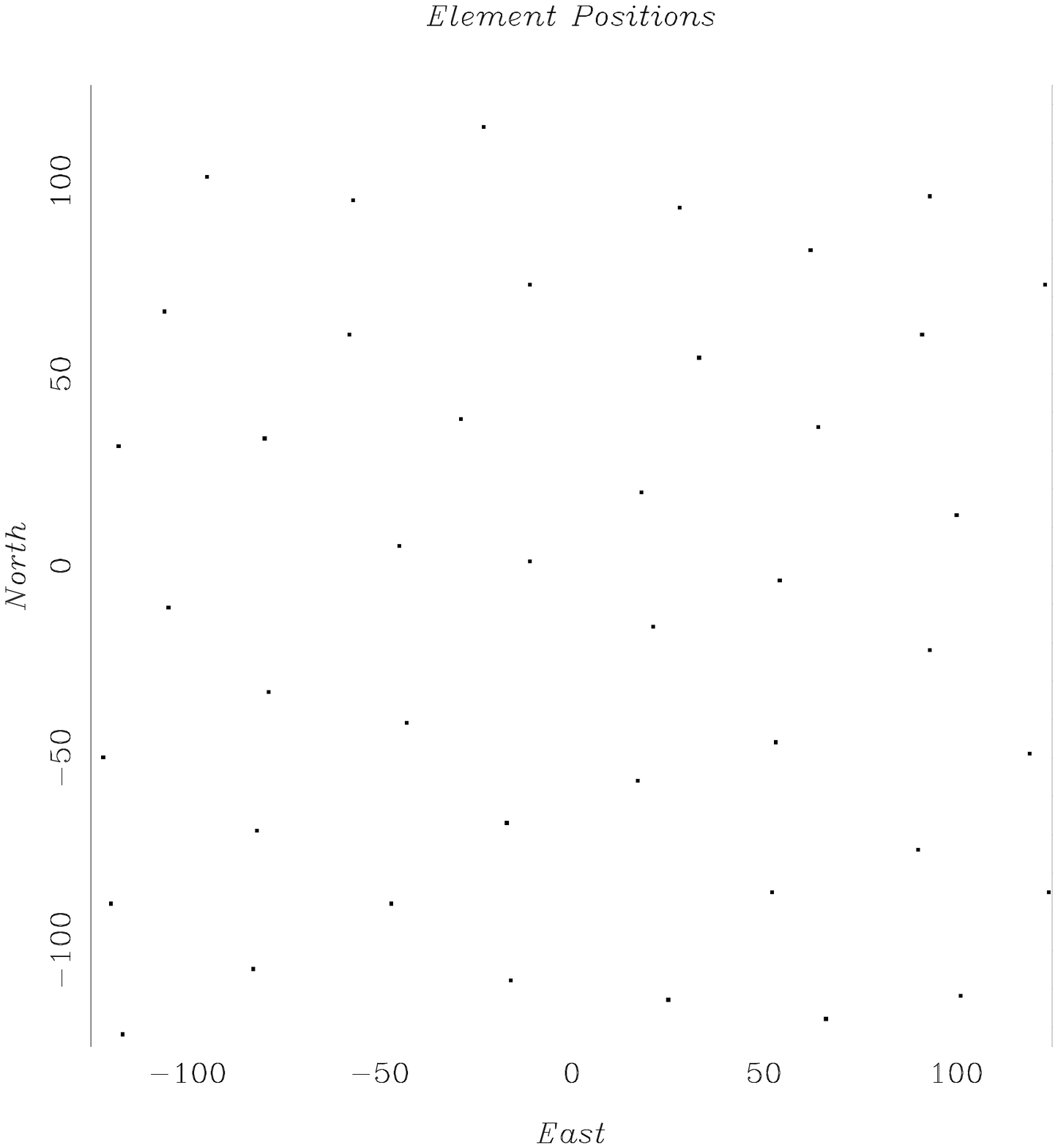}\includegraphics[width=6.75cm]{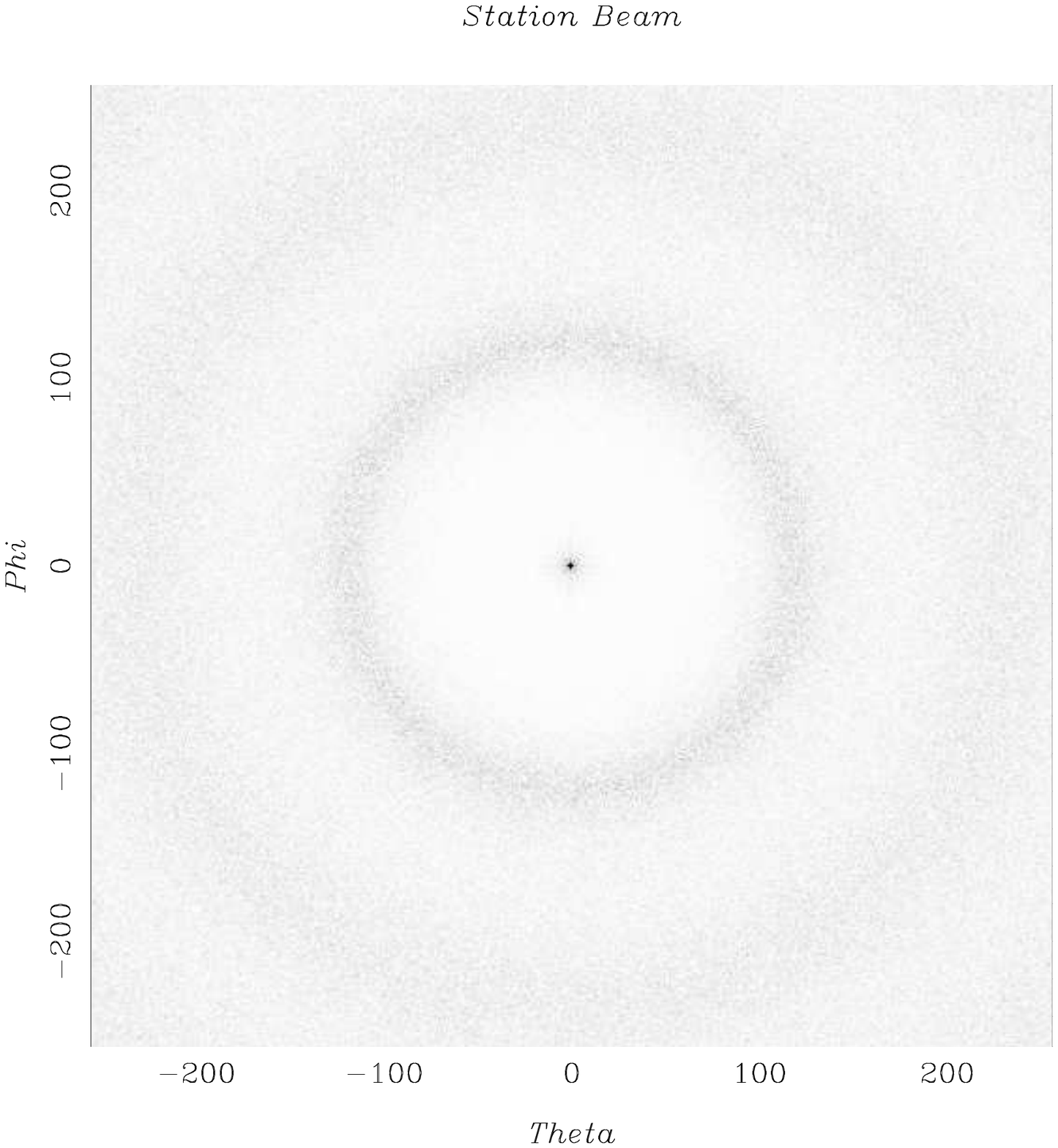}\quad\includegraphics[width=6cm]{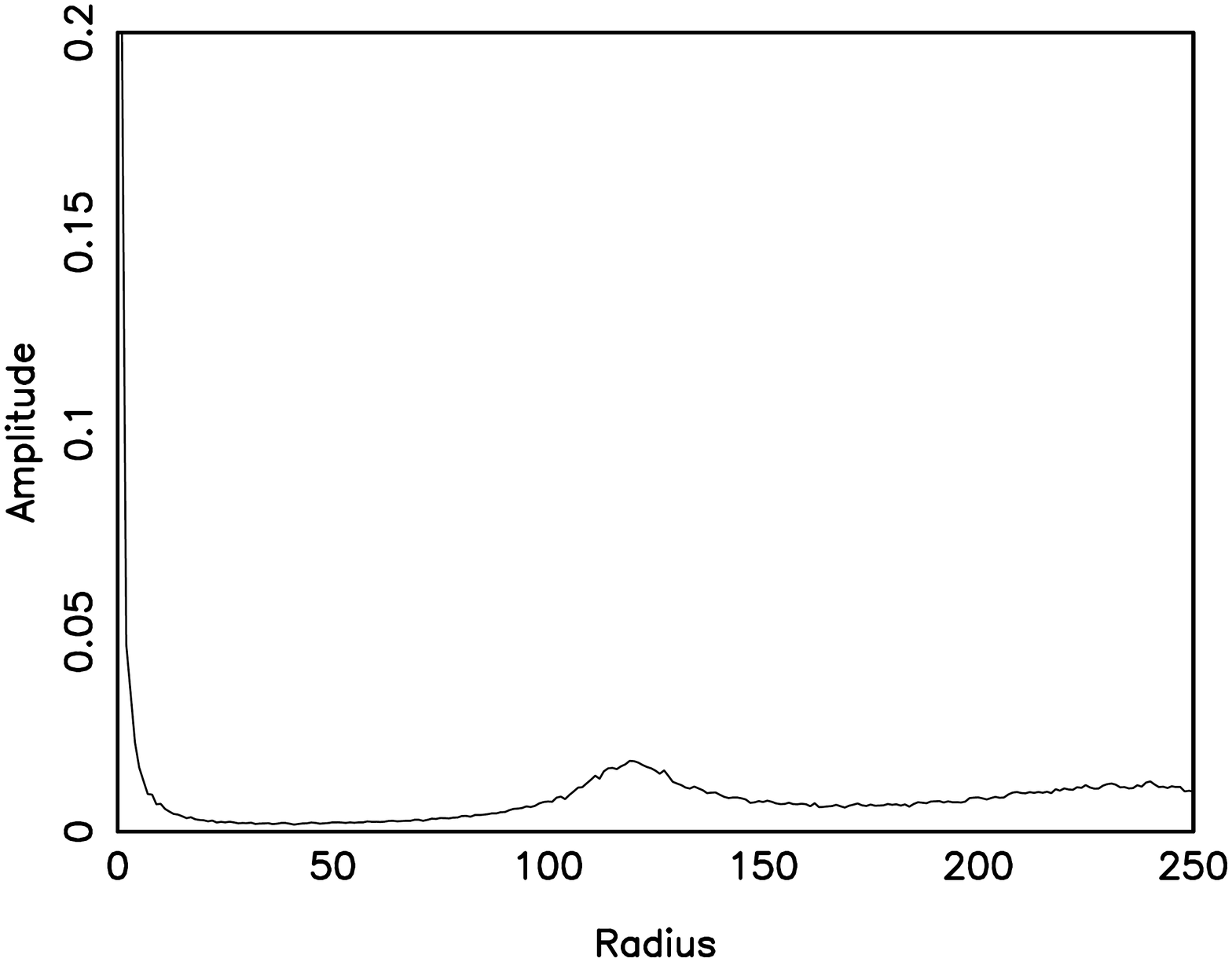}
\caption{\label{hex_rnd}From left: element configuration; station
  beam; radial station beam power. Top plot show an infinite regular
  configuration of delta function elements; bottom plots show a
  randomised configuration with the same mean~inter-element~spacing.}
\end{figure}

\end{landscape}

Consider an aperture array of infinite extent, comprised of
mono-chromatic elements with delta function receiving area, configured
in a hexagonal close packed arrangement and beam formed with equal
weights and zero phase offsets. The Fourier transform of this
configuration is the voltage station beam of the an array and will be
a set off hexagonally close packed delta functions of equal amplitude
distributed over an infinite extent (see Figure~\ref{hex_rnd},
top. Note that these illustrative figures are simulated with 9000
rather than an infinite number of elements and some non-ideal features
are generated). This can be thought of as a main beam pointing towards the
zenith and then sidelobes (hereafter ``far-out sidelobes'' for reasons
that will become apparent) arranged in a hexagonal pattern around
this. If one reduces the distance (in wavelengths) between elements
the distance between the sidelobes of the beam pattern increases, and
vice versa.

\begin{landscape}

\begin{figure}[H]
\includegraphics[width=6.75cm]{./rnd_inf_stn}\includegraphics[width=6.75cm]{./rnd_inf_pb}\quad\includegraphics[width=6cm]{./rnd_inf_pt}\\
\includegraphics[width=6.75cm]{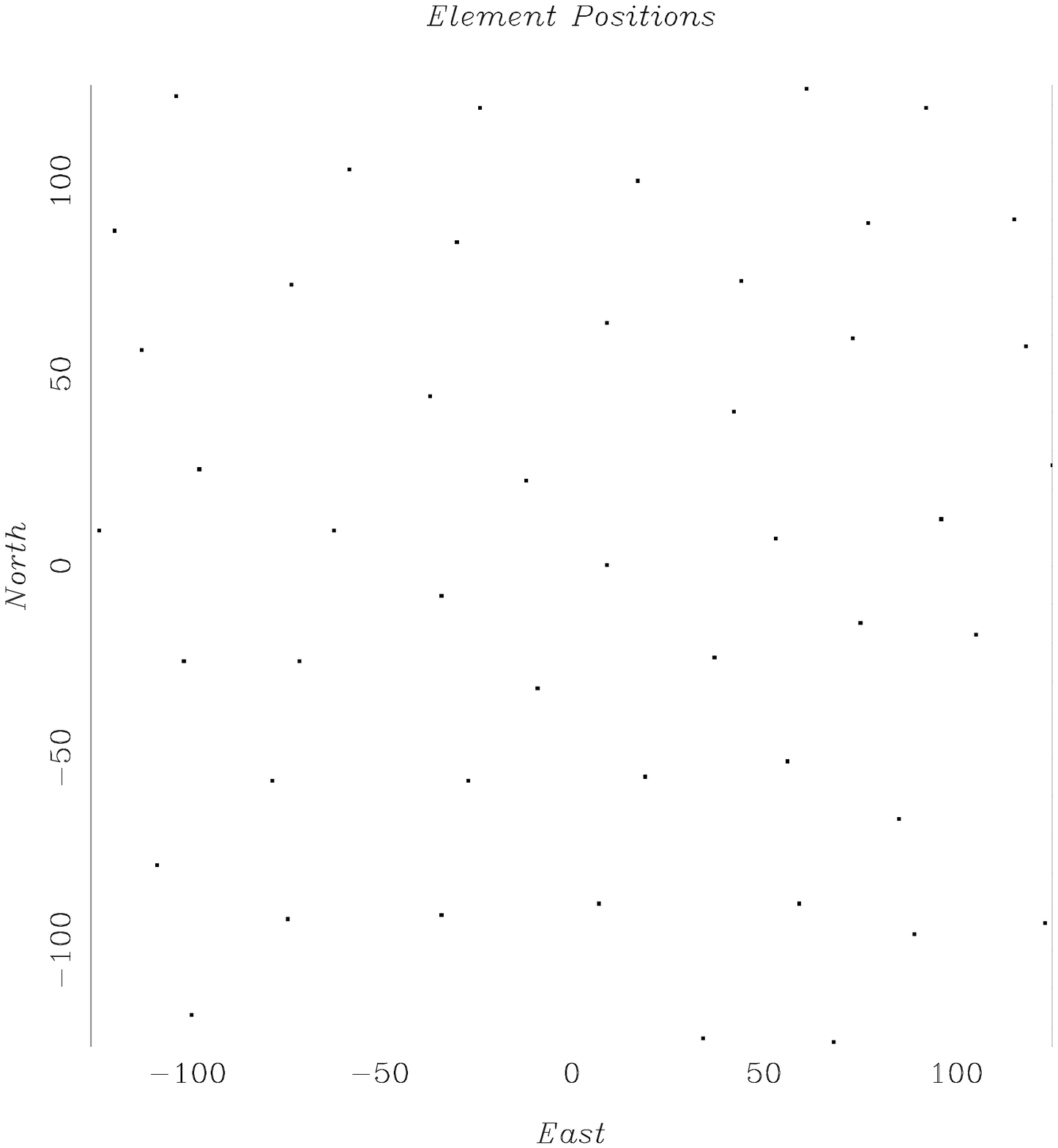}\includegraphics[width=6.75cm]{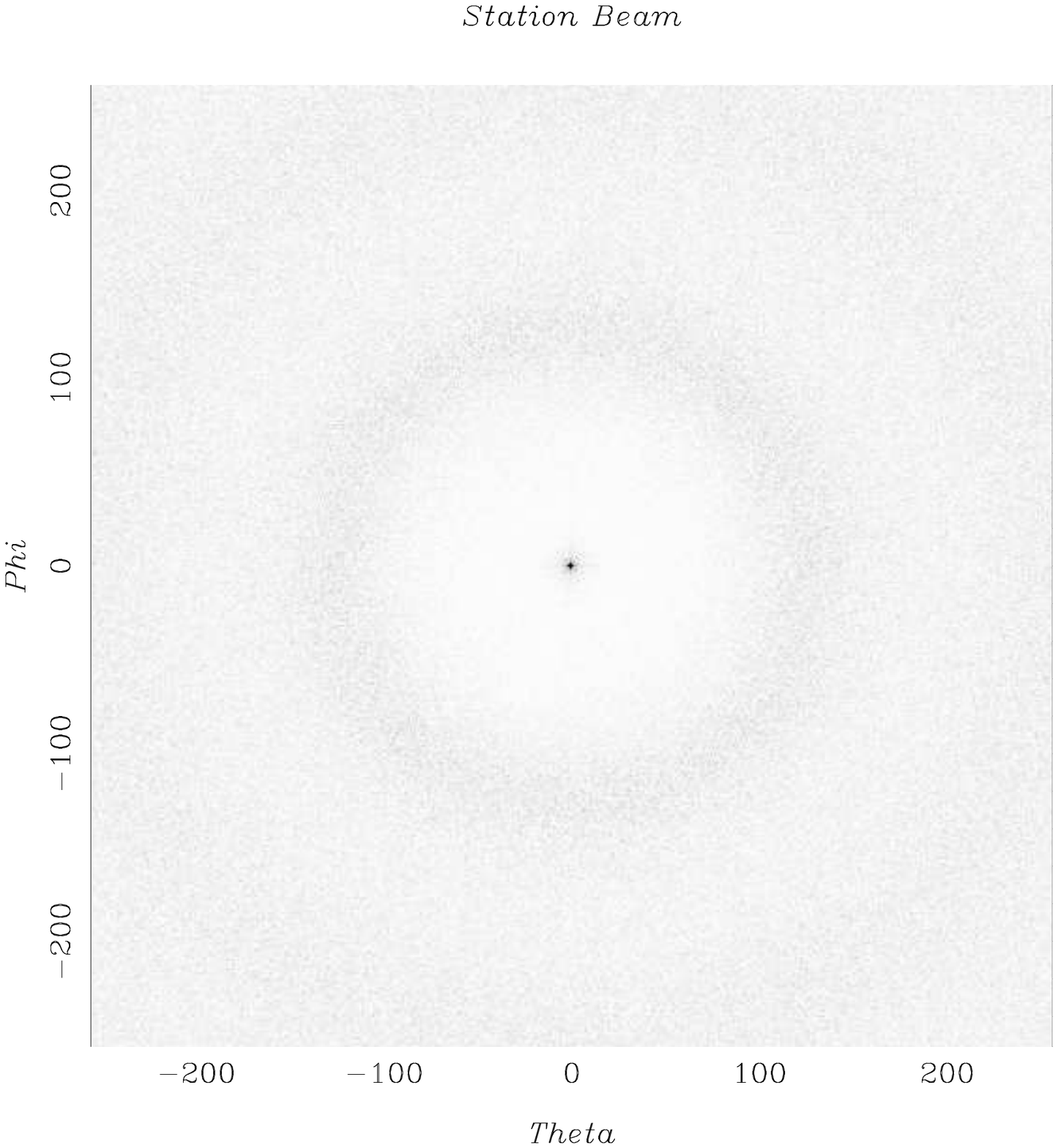}\quad\includegraphics[width=6cm]{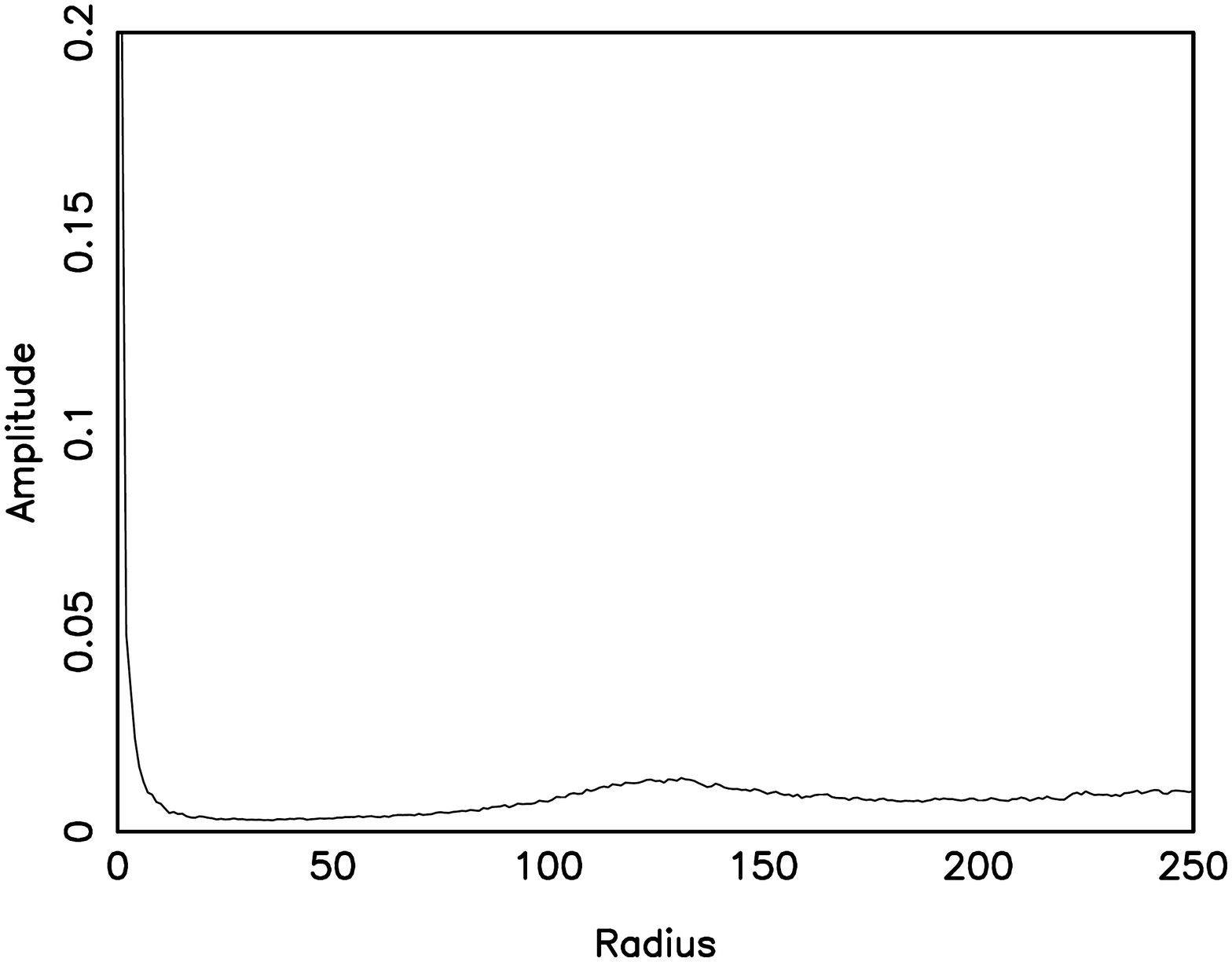}
\caption{\label{diff_minspac}Plots as in Figure~\ref{hex_rnd}. Top
  plots show results for randomised configuration where minimum
  spacing is 85\% of the mean spacing; bottom plots where it is 75\%
  with the same mean. Note that the sidelobe falls in the same place
  but is broadened.}
\end{figure}

\end{landscape}

If one now randomises the configuration while maintaining a minimum
spacing between elements the far-out sidelobes become a circularised
band with amplitude that is lower than that of the zenith beam and
that falls off going to longer radius (see Figure~\ref{hex_rnd},
bottom). The radii of these bands from the main lobe is determined by
the mean inter-element spacing and the thickness of the band by the
distribution of distances to nearest element neighbours (see
Figure~\ref{diff_minspac}). The main beam is still a delta function
(this is not quite apparent in the plots since a finite simulation was
used).

Replacing the delta function receiving area of the elements with a
Gaussian is equivalent to convolving the station illumination function
with the Gaussian. The effect on the Fourier transform i.e. the
station beam is therefore to multiply by a Gaussian (see
Figure~\ref{delta_gaus}). This can be though of as incorporating the
angular response pattern of the element, which suppresses the
sidelobes.

Finally, one can produce a realistic station configuration with a
finite number of elements by multiplying the infinite extent array by
a finite function e.g. a circular tophat. The effect on the station
beam is therefore to convolve with the Fourier transform of this
finite function. For a uniformly weighted circular station, this will
result in the main beam delta function being replaced by a function
with strong, ``near-in'' sidelobes (see Figure~\ref{finite}, top). If
apodisation is applied, these near-in sidelobes can be suppressed, as
discussed in Section~\ref{shared} (also see Figure~\ref{finite},
bottom). The far-out sidelobes are also convolved with this same
function, but this will not lead to any significant suppression. Note
that the other, independent, effect of reducing the number of elements
within a station is to increase the strength of the far sidelobes.

These near-in and far-out sidelobes therefore have different physical
origins. The near-in sidelobes are due to the station size, $D$ and
therefore appear at an angular scale size proportional to
$D/\lambda$. The far-out sidelobes are due to the fact that the
station is comprised of individual elements, mean spacing $d$, and
therefore appear at an angular scale size proportional to
$d/\lambda$. Since $D \approx d \sqrt{N}$, where $N$ is the number of
elements per station, then the number of near-in sidelobes before the
first far-out sidelobe is proportional to $\sqrt{N}$. Clavier et
al. \cite{clavier} present a more sophisticated analysis of this
phenomenon (using the terms ``aperture-type'' and ``non-coherent''
sidelobes) and find that the number of sidelobes inside the far-out
sidelobes is well approximated by $0.3 \sqrt{N}$. They also find that
the power in the far sidelobes goes down as $N$.

Therefore implementing an apodisation to the station illumination can
significantly reduce near-in sidelobes, but to first order this will
not affect far-out sidelobes. These far-out sidelobes must therefore
be suppressed: by the intrinsic element beam pattern; or by time and
frequency averaging; or by increasing the number of elements in each
station.

\begin{landscape}

\begin{figure}[H]
\includegraphics[width=6.75cm]{./rnd_inf_stn}\includegraphics[width=6.75cm]{./rnd_inf_pb}\quad\includegraphics[width=6cm]{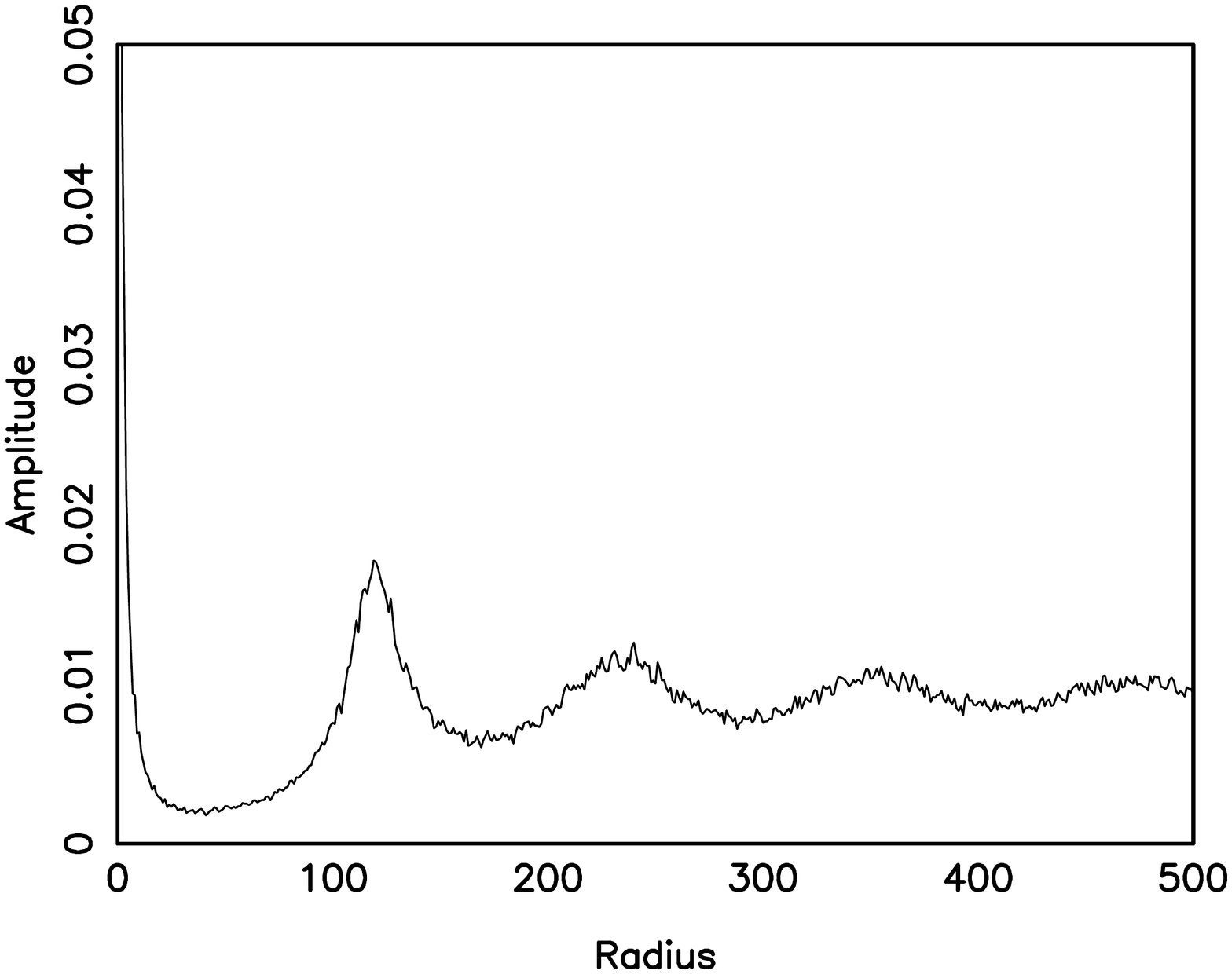}\\
\includegraphics[width=6.75cm]{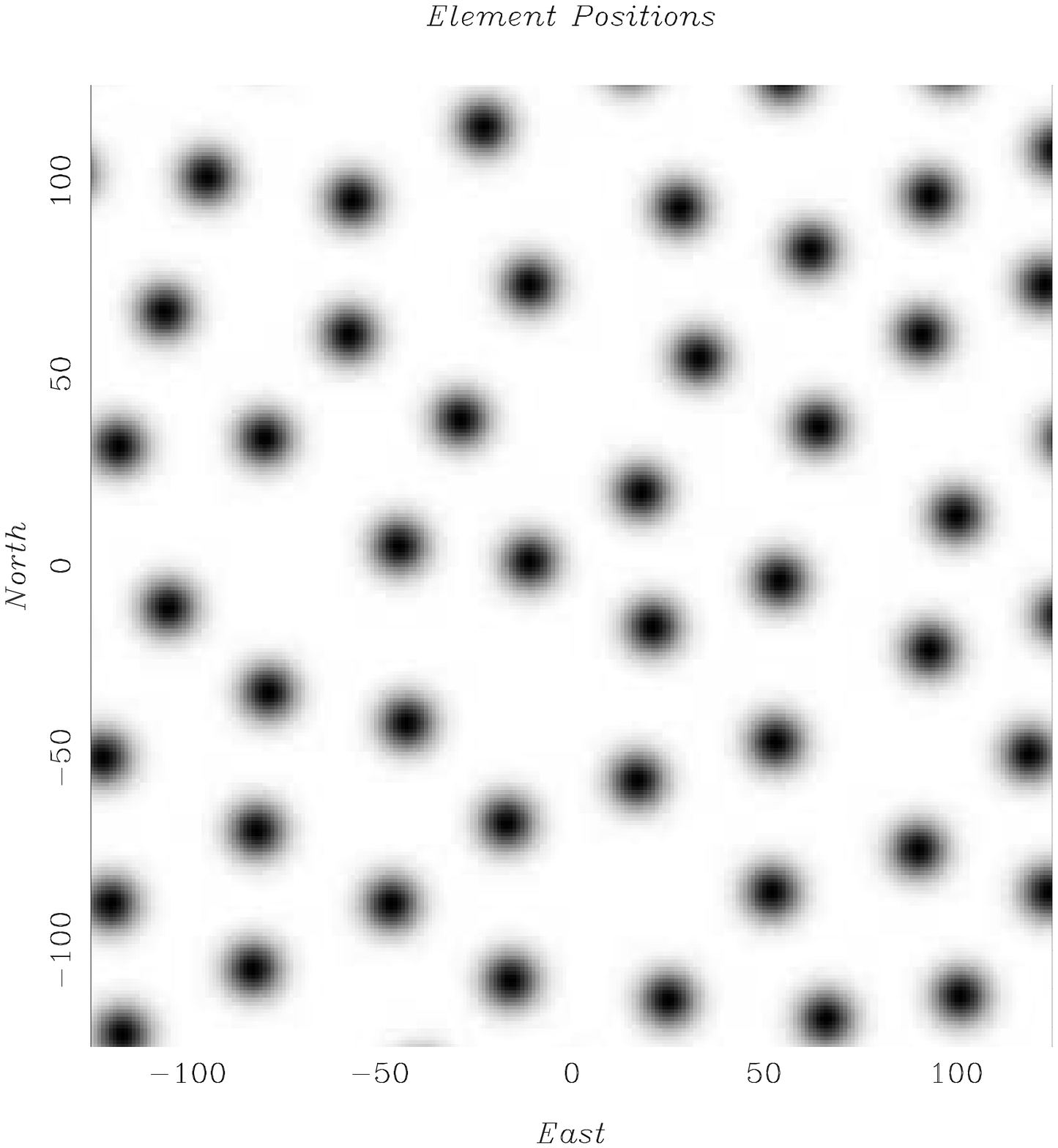}\includegraphics[width=6.75cm]{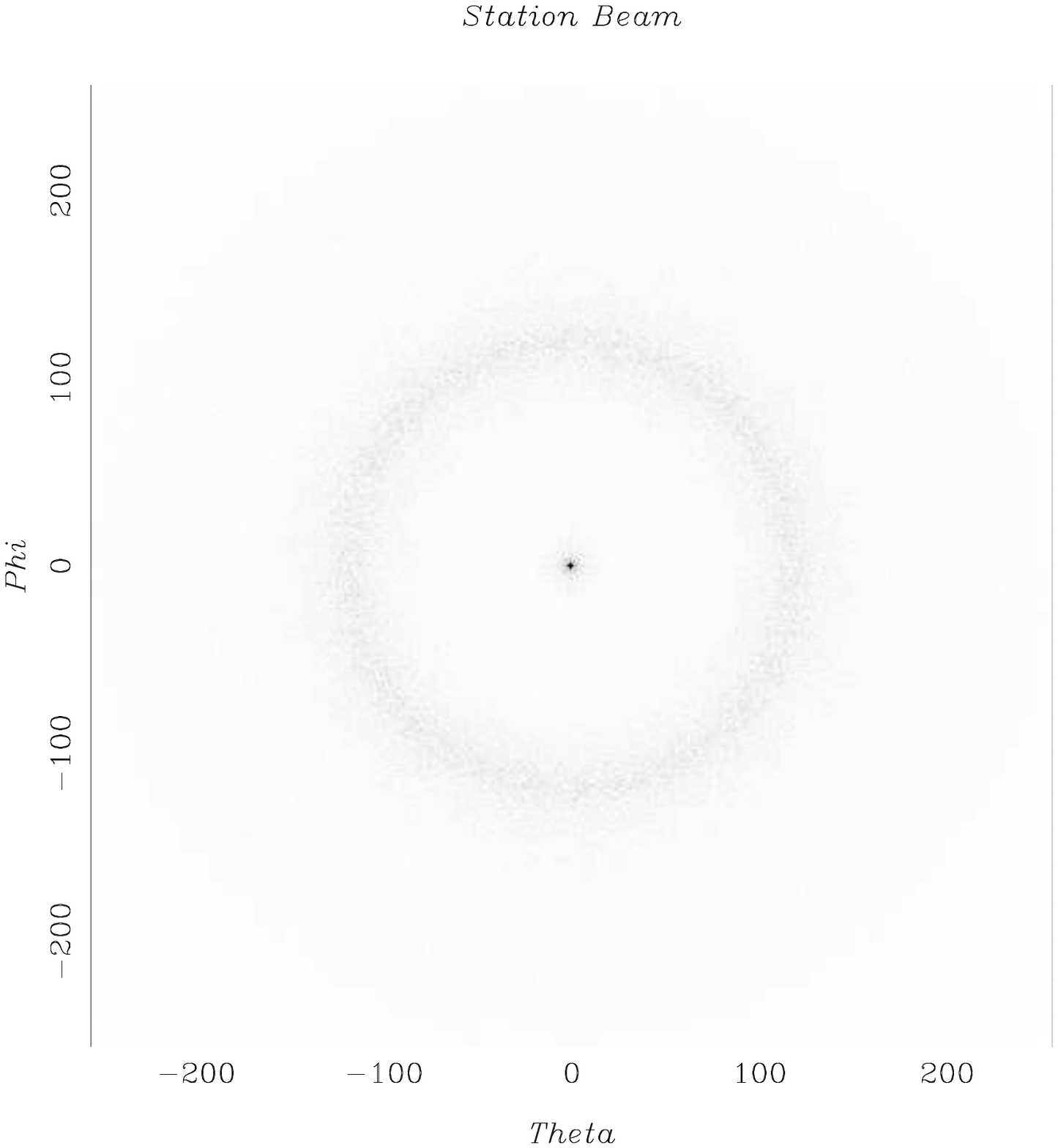}\quad\includegraphics[width=6cm]{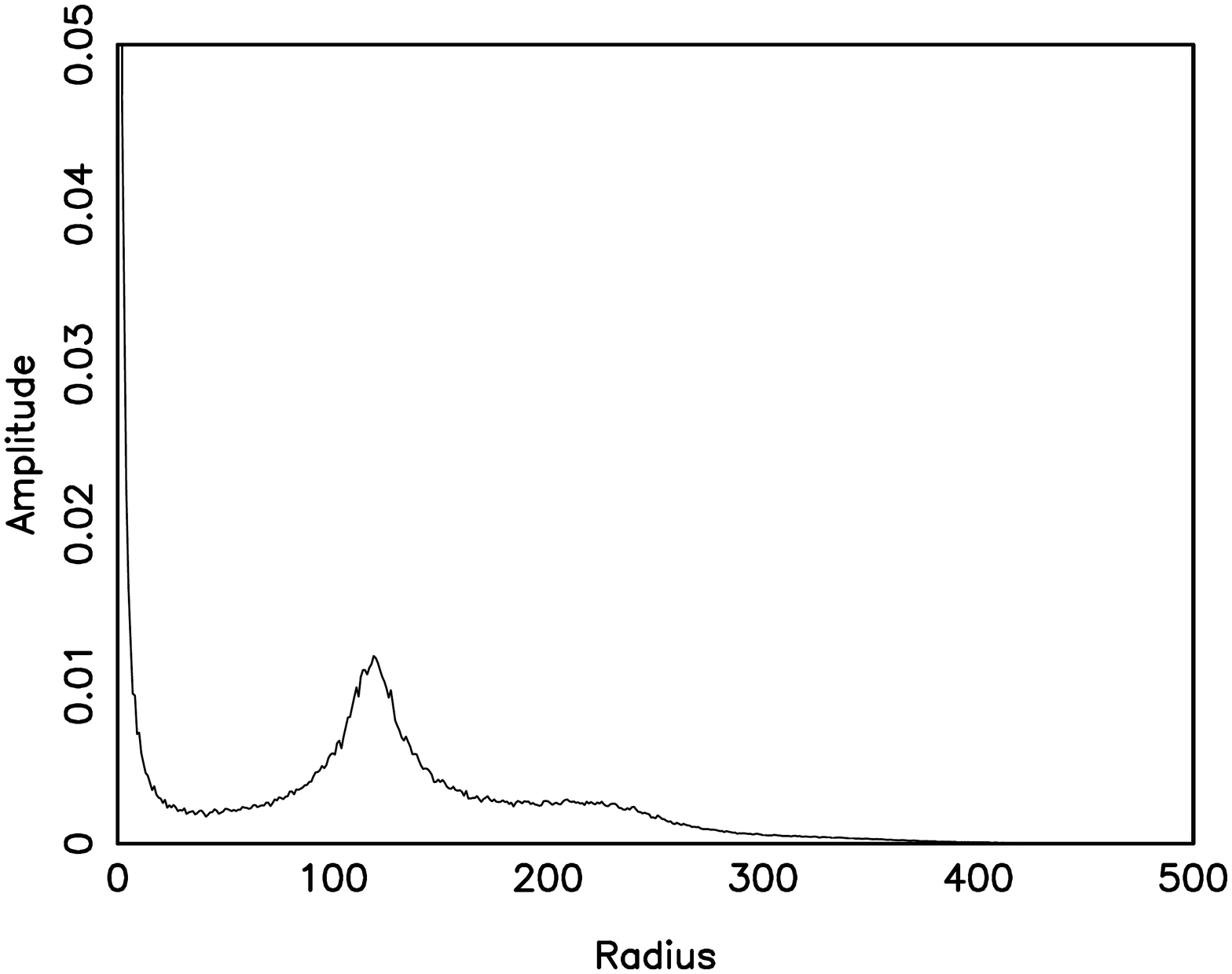}
\caption{\label{delta_gaus}Plots as in Figure~\ref{hex_rnd}. Plots
  show the effect of moving from delta function response elements to a
  Gaussian response. Note that the sidelobes are suppressed.}
\end{figure}

\begin{figure}[H]
\includegraphics[width=6.75cm]{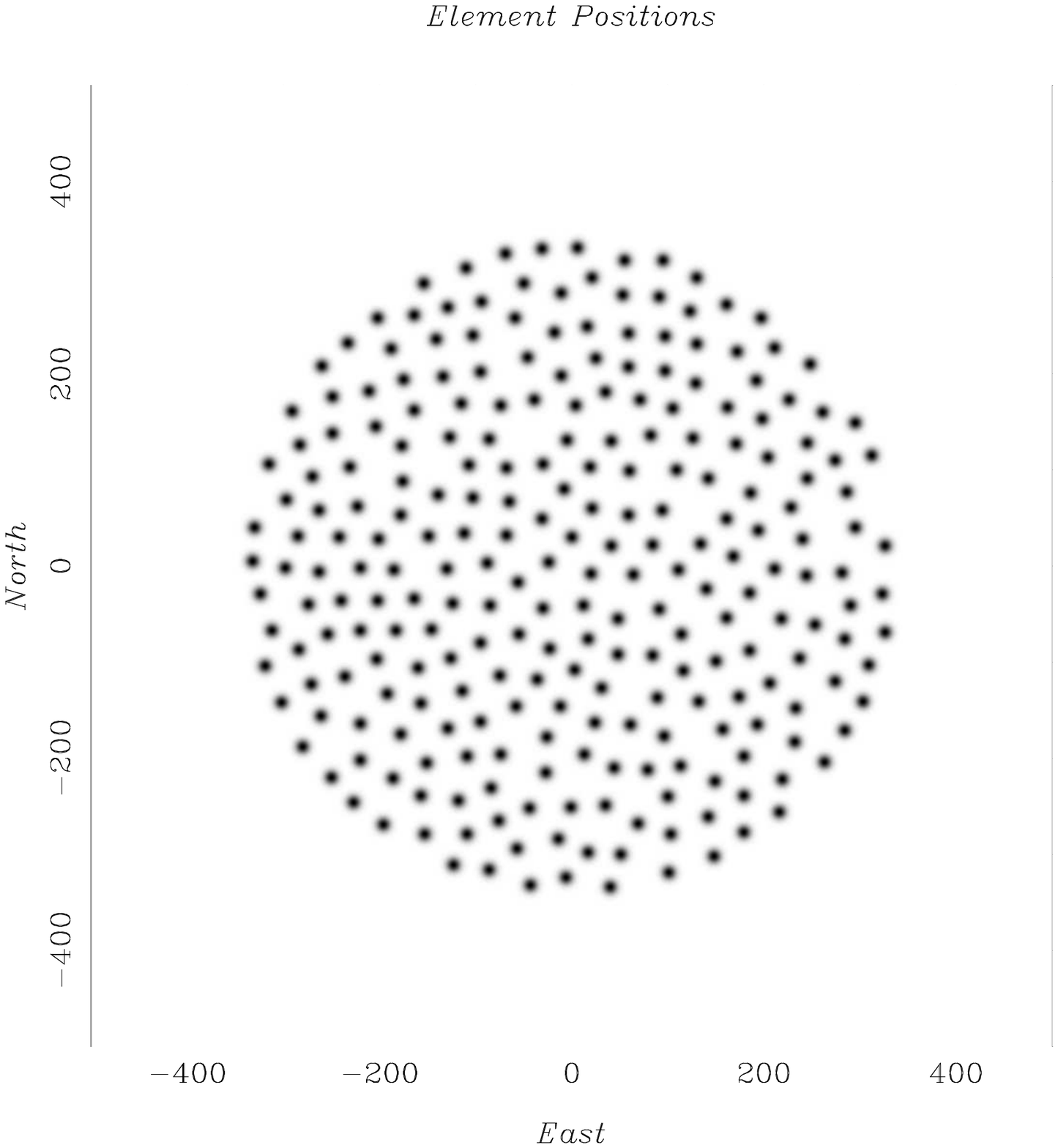}\includegraphics[width=6.75cm]{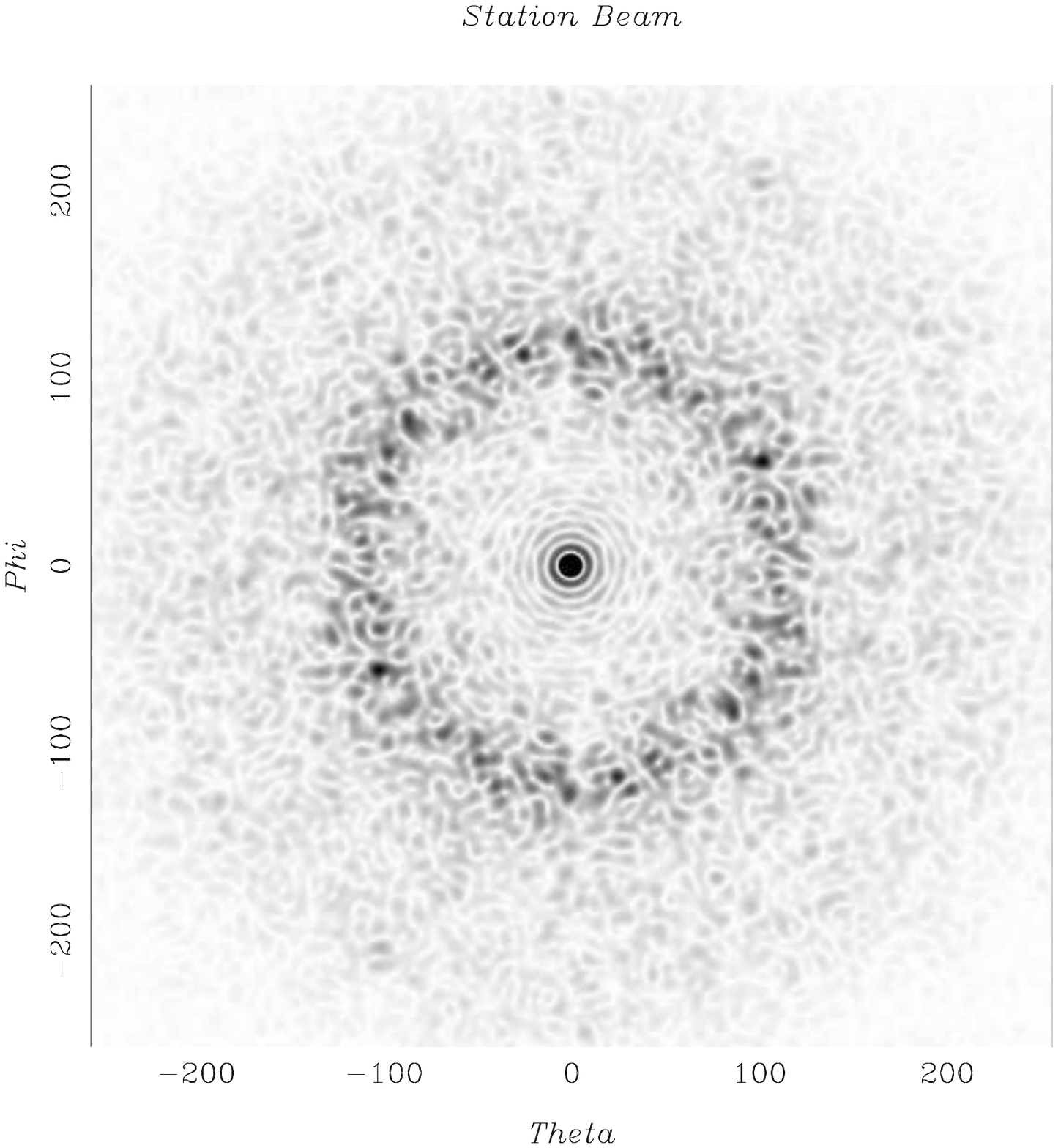}\quad\includegraphics[width=6cm]{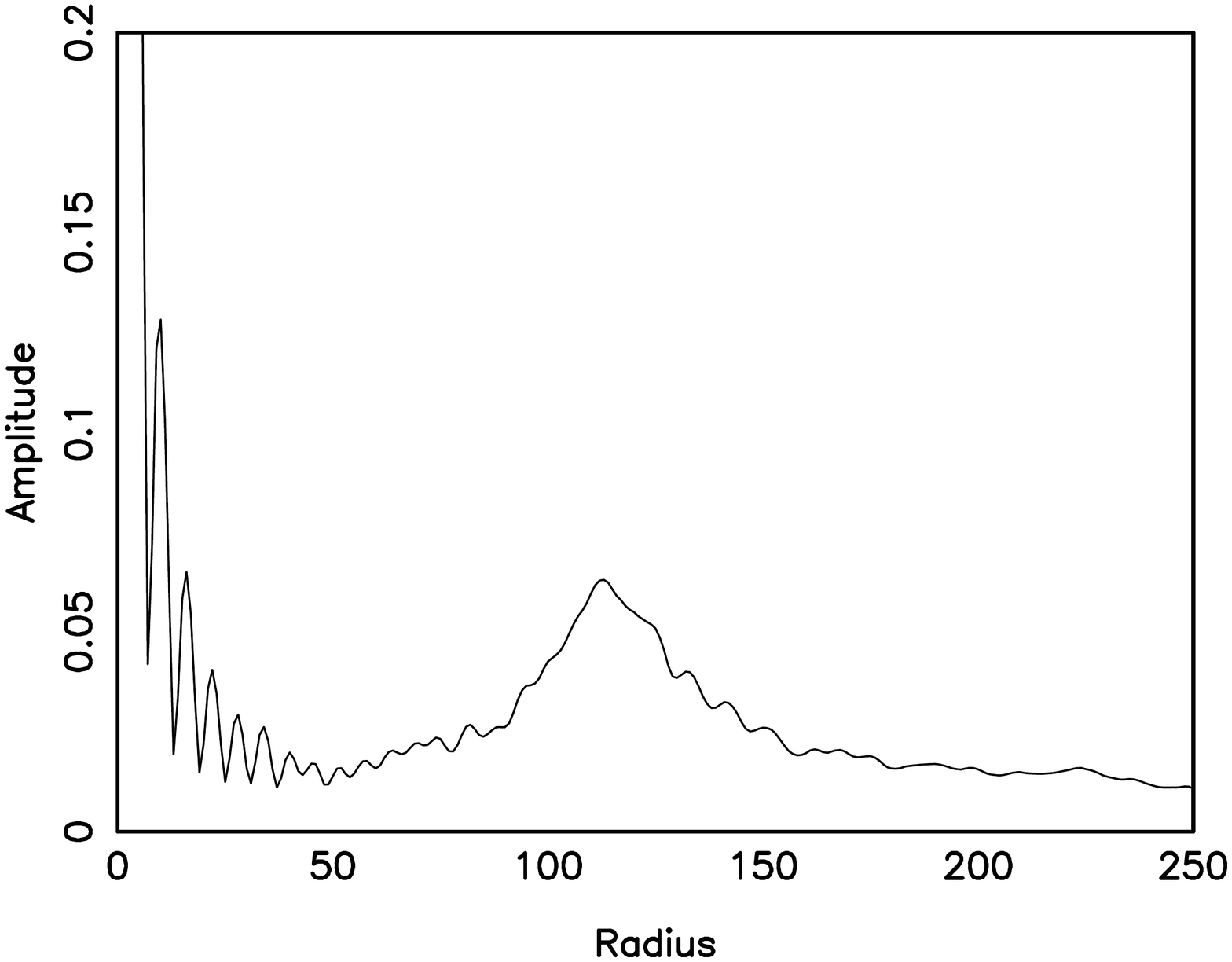}\\
\includegraphics[width=6.75cm]{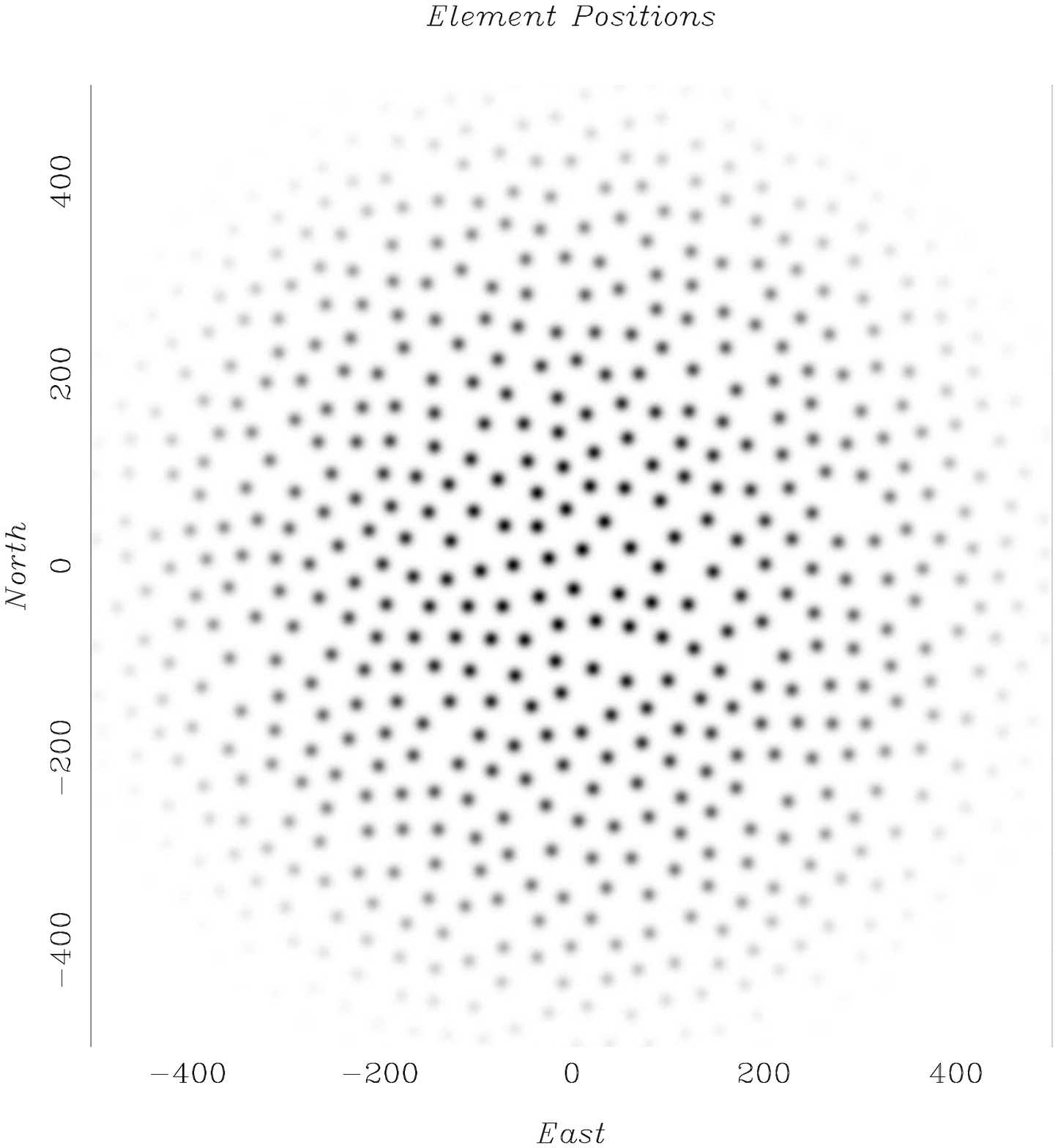}\includegraphics[width=6.75cm]{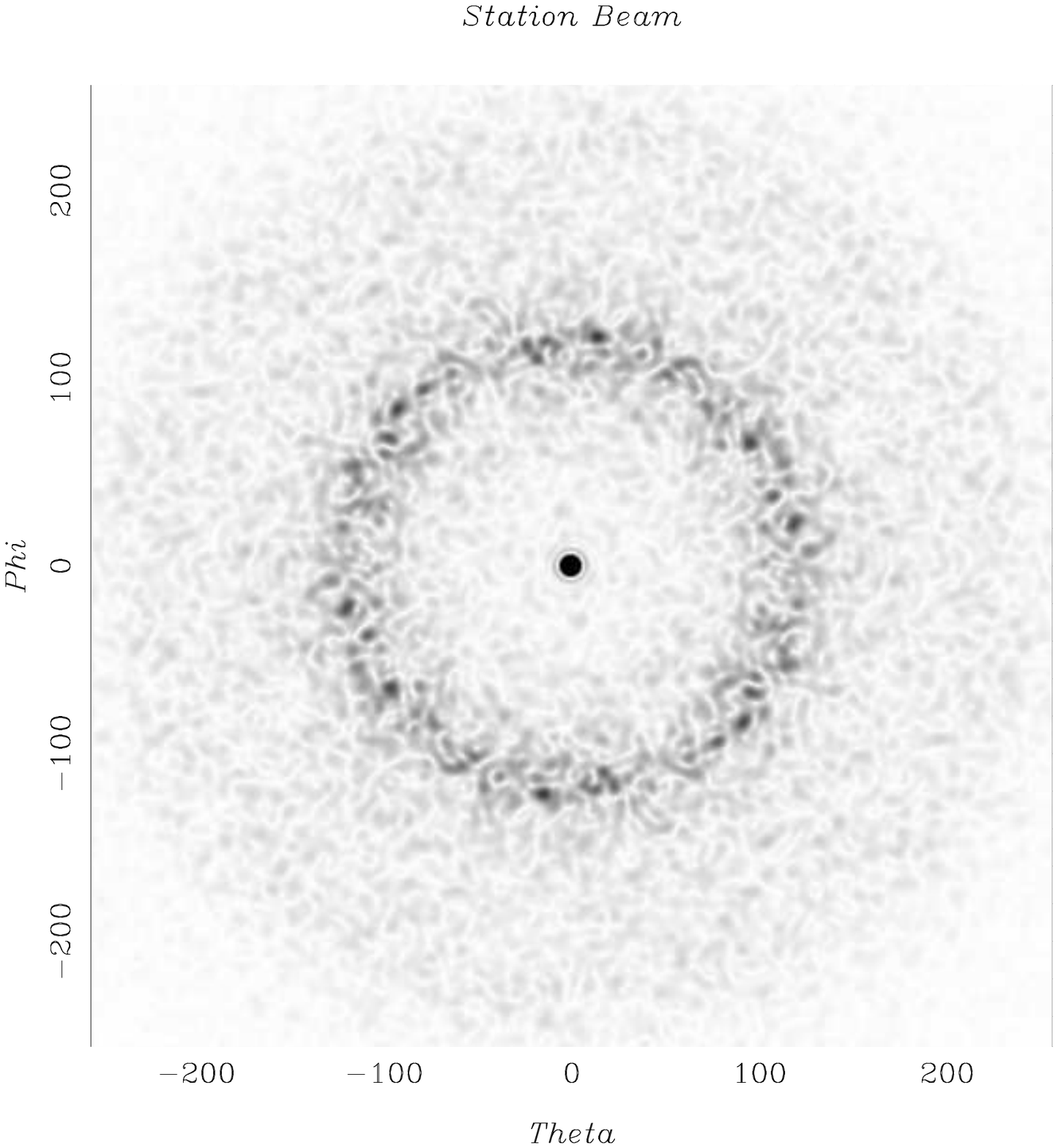}\quad\includegraphics[width=6cm]{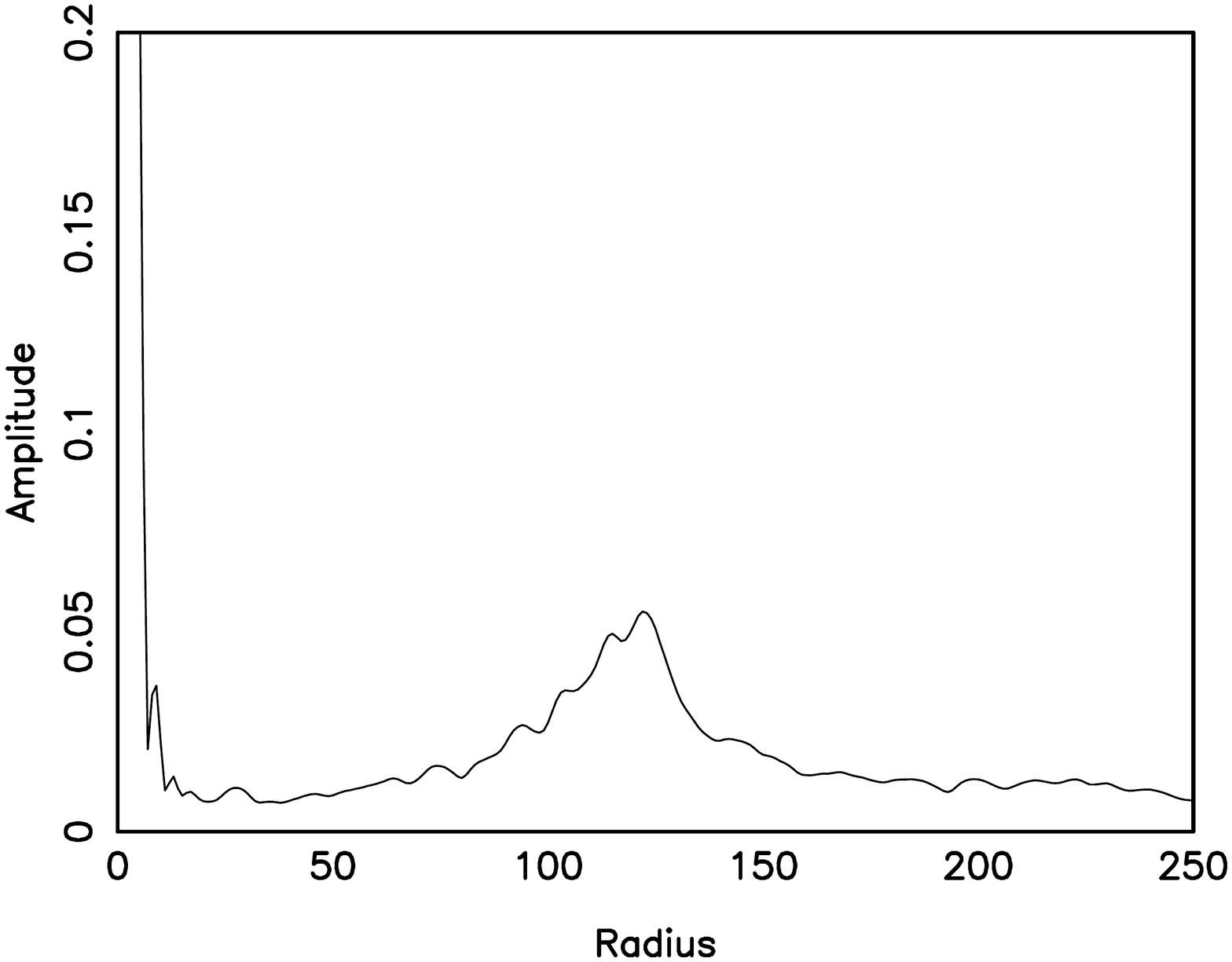}
\caption{\label{finite}Plots as in Figure~\ref{hex_rnd}. The top plot
  show the effect of moving to a finite sized station; note that the
  inner beam is now convolved with a function giving strong
  sidelobes. The bottom plot shows a finite but apodised station which
  has much suppressed near-in sidelobes.}
\end{figure}

\end{landscape}

\section{Station configuration within the core}\label{stn_conf}

\begin{figure}[H]
\begin{center}
\includegraphics[width=8cm,angle=0]{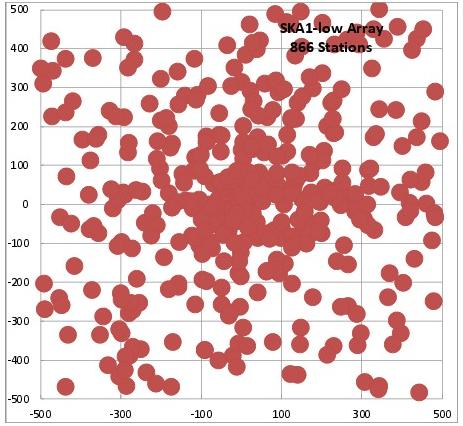}
\caption{\label{bd} Configuration proposed in the Baseline Design
  (Dewdney et al.) with a fully filled central region and then a
  random distribution of stations with a decreasing density.}
\end{center}
\end{figure}

The Baseline Design envisages that the core of SKA-low will be close
to miximally packed within the inner 250m radius with the density of
stations then dropping off as a Gaussian. There will be 374 stations
with 500m and 647 within 1km. Station positions are randomised within
this area. This gives a configuration such as shown in
Figure~\ref{bd}. Here I consider a different configuration within the
inner 1km radius which gives the same density of stations as a
function of radius, but has some features which may give important
advantages.

\begin{figure}[H]
\begin{center}
\includegraphics[width=6cm,angle=270]{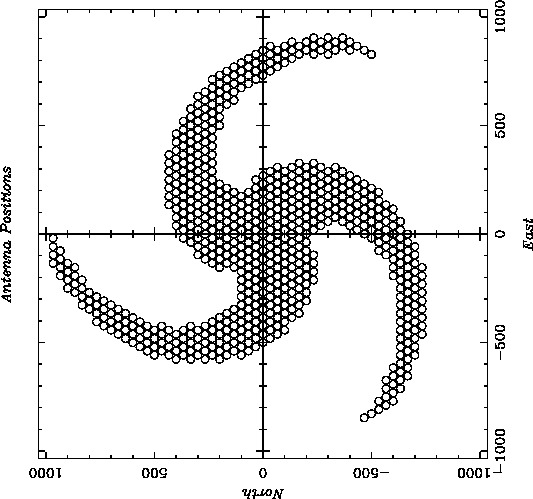}\quad\includegraphics[width=6cm,angle=270]{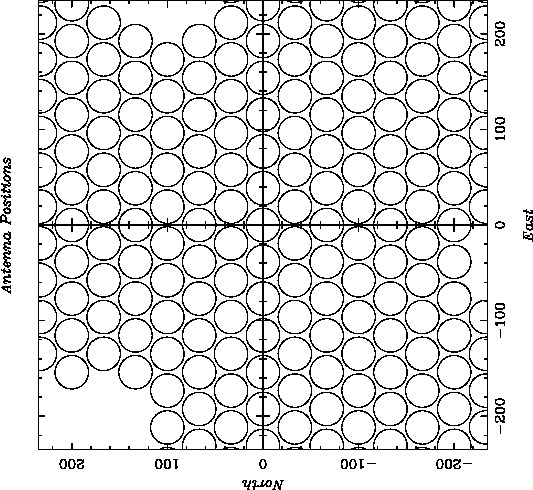}
\caption{\label{reg_conf} Alternative proposed configuration where
  the reduction in radial station density is implemented though spiral
  arms, but within which the elements are fully filled. Station
  positions have are based on a regular hexagonal close packed
  grid. The righthand plot shows a close up of the inner region}
\end{center}
\end{figure}

The new configuration is again fully filled in the central 200m but
then implements the density drop with radius with fully filled spiral
arms (whose thickness drops with radius) as shown in
Figures~\ref{rand_conf} and~\ref{reg_conf}; in the former the station
positions have undergone randomisation while in the latter they are
based on a regular grid. The difference between these two variants are
discussed later. The configuration has been tailored to give
approximately the desired number of stations within each radial annulus
as shown in Figure~\ref{cumul} These configurations offer the
following possible advantages:

\begin{figure}[H]
\begin{center}
\includegraphics[width=6cm,angle=270]{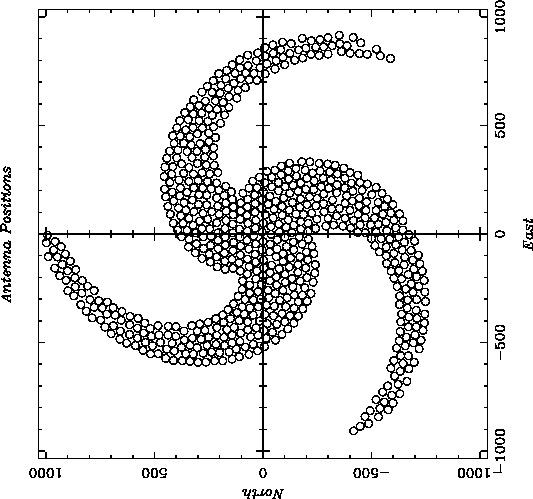}\quad\includegraphics[width=6cm,angle=270]{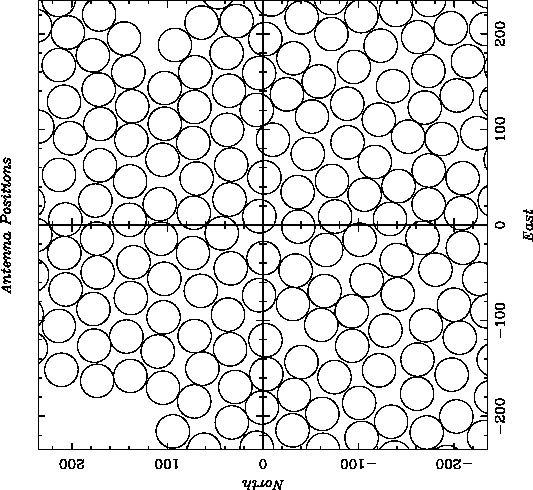}
\caption{\label{rand_conf} Alternative proposed configuration where
  the reduction in radial station density is implemented though spiral
  arms, but within which the elements are fully filled. Station
  positions have been randomised. The righthand plot shows a close up
  of the inner region.}
\end{center}
\end{figure}

\begin{figure}[H]
\begin{center}
\includegraphics[width=6cm,angle=270]{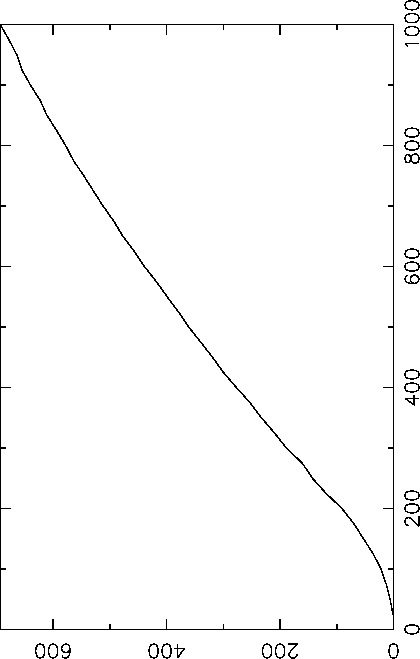}
\caption{\label{cumul} Cumulative number of stations against radius
  for the configuration shown in Figure~\ref{rand_conf}. There are 361
  stations within 500m radius and 691 stations within 1km.}
\end{center}
\end{figure}

\begin{figure}[H]
\begin{center}
\includegraphics[width=6cm,angle=270]{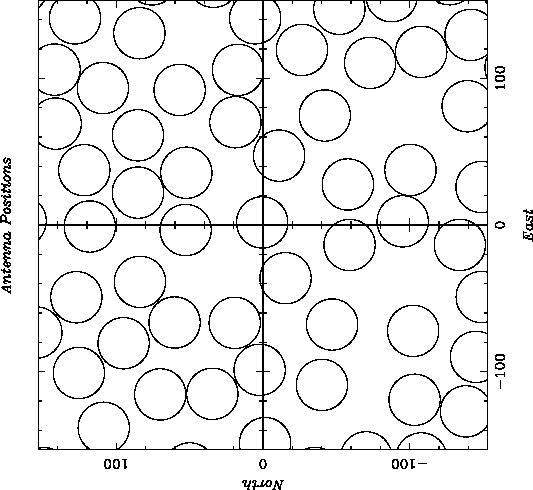}
\caption{\label{sparce} A realisation of Baseline Design configuration
  just outside the fully filled region with a filling factor of
  approximately 2 lower than fully filled.  Increasing the density of
  the AA for SKA2 will cause significant logistical problems. In
  addition, it will be difficult to identify possible sites for new
  35m stations.}
\end{center}
\end{figure}

\begin{figure}[H]
\begin{center}
\includegraphics[width=8cm,angle=270]{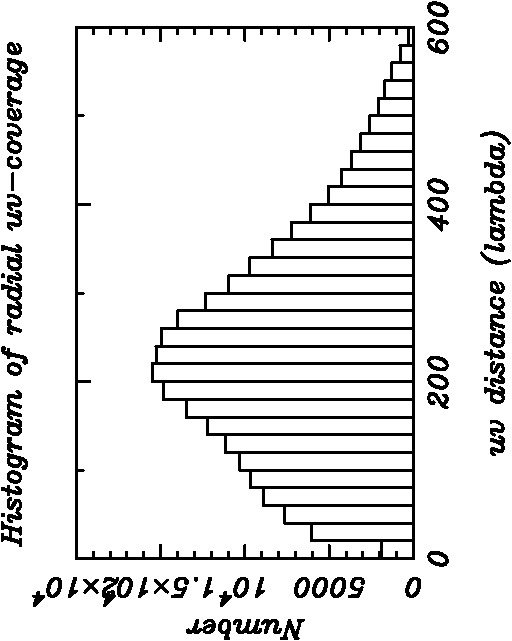}
\caption{\label{histogram}Histogram of number of baselines (measured
  in lambda assuming an observing frequency of 100 MHz) generated by
  the configuration shown in Figure~\ref{rand_conf}.}
\end{center}
\end{figure}

\begin{itemize} 

\item Expansion to SKA2. In SKA2 the density of stations with radius
  will be increased dramatically and the fully filled region of the
  core will be expanded. With the baseline design configuration shown
  in Figure~\ref{bd}, accessing the site for a new station, bringing
  in equipment and installing it will be challenging amongst the
  existing stations. Adding to the configuration shown in
  Figure~\ref{rand_conf} will be considerably easier. In addition,
  just beyond the fully filled region of the Phase 1 core the density
  of stations will be so high that it will be close to impossible to
  find a 35m diameter site in which to locate a new station; this is
  shown in Figure~\ref{sparce}. Finally, access for maintenance will be 
  considerably easier with this spiral arm configuration.

\item Variable sized stations. The proposed design from the AA
  consortium~\cite{LFAA} envisages a flexible approach to station
  size, with possible diameters ranging from 20m to 100m. This
  approach would fail beyond the fully filled region in the Baseline
  Design configuration since station sizes are defined when they are
  deployed --- there are no more antenna beyond the station edge to
  allow for the desired flexible approach. In the proposed
  configuration the spiral arms are fully filled and so allow for the
  possibility for redefinition of stations dynamically.

\item Overlaping, apodised stations. As described in Section~\ref{shared}
  document there are many potential benefits to the station beam shape
  and calibratability that could be realised through defining
  overlapping stations where a radial weighting is applied during
  beamforming to suppress far sidelobes. Similarly to above, this does
  require a fully filled configuration such as would be provided by
  the proposed filled spiral arms.

\item Number of short baselines. Since the spiral arms are fully
  filled they will generate a reasonable number of shortest possible
  baselines which are critical for brightness sensitivity on the
  largest angular scales, see Figure~\ref{histogram}. This is
  important for the EoR experiment.

\end{itemize}

\begin{figure}[H]
\begin{center}
\includegraphics[width=6cm,angle=270]{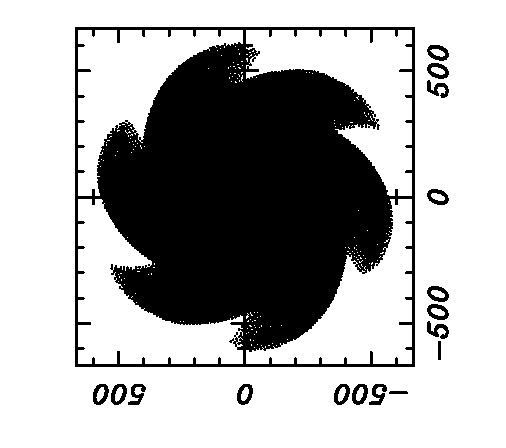}\quad\includegraphics[width=6cm,angle=270]{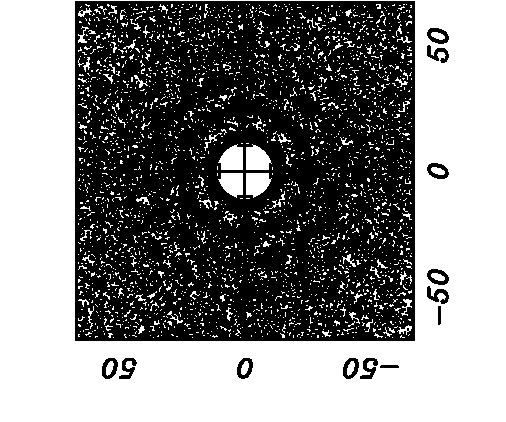}
\caption{\label{uv_cov} Snapshot coverage of the uv-plane for the
  configuration shown in Figure~\ref{rand_conf} assuming an observing
  frequency of 100 MHz. The righthand plot shows a close up view of
  the centre of the aperture plane.}
\end{center}
\end{figure}

\begin{figure}[H]
\begin{center}
\includegraphics[width=6cm,angle=270]{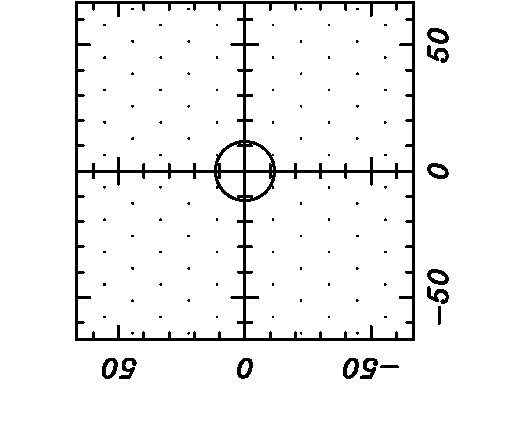}
\caption{\label{uv_inner_reg} Snapshot coverage of the uv-plane for
  the configuration shown in the righthand plot of
  Figure~\ref{rand_conf} assuming an observing frequency of 100
  MHz. The regular grid adopted results in very high redundancy of
  baselines and so a correspondingly poor coverage of the
  uv-plane. The circle radius approx 10 lambda is the size of the
  aperture illumination function. }
\end{center}
\end{figure}

The main disadvantage to the proposed configuration is that the
uv-coverage will necessarily be less uniform that that generated by
the Baseline Design configuration. However, since there are 647
station the coverage can still be excellent as long the configuration
is randomised. Figure~\ref{uv_cov} shows the snapshot uv-plane
coverage corresponding to Figure~\ref{rand_conf}, with a close up of
the inner region. The uv-coverage therefore looks very good, but could
of course be improved further with earth-rotation synthesis. For
comparison, Figure~\ref{uv_inner_reg} shows the inner part of the
uv-plane for the un-randomised configuration shown in
Figure~\ref{reg_conf}.

\section*{Acknowledgements}

I would like to thank Nima Razavi-Ghods, Eloy de Lera Acedo, Benjamin
Mort, Fred Dulwich, Paul Alexander, Andrew Faulkner and Stef Salvini
for useful conversations.


\begin{thebibliography}{50}
\bibitem{braun} Braun R., 2013, A\&A 551, A91

\bibitem{clavier} Clavier, T. et al., 2013, IEEE Transactions on
  Antennas and Propagation, 99, 1

\bibitem{BD} Dewdeney, P.E., 2013, SKA-1 System Baseline design,
  SKA-TEL-SKO-DD-001-1\_BaselineDesign1.

\bibitem{LFAA} Faulkner, A., 2013, Low Frequency Aperture Array 
 Technical Description, AADC-TEL.LFAA.SE.MGT-AADC-PL-002

\bibitem{gonzalez-ovejero} Gonzalez-Ovejero, D., 2011, IEEE
  International Symposium on Antennas and Propagation, 1762

\bibitem{eloy} de Lera Acedo, E., 2011, IEEE Transactions on Antennas
  and Propagation, vol. 59, 1808

\bibitem{eloy2} de Lera Acedo, E., 2012, International Conference on
  Electromagnetics in Advanced Applications, 353.

\bibitem{eloy3} de Lera Acedo, E., et al.,  2011 International Conference on 
Electromagnetics in Advanced Applications, 390.

\bibitem{stefan} Wijnholds, S. J., Bregman J. D. and van Ardenne, A.,
  2011, Radio Science, 46, RS0F07

\end{thebibliography}
\end{document}